\documentclass[10pt,journal,cspaper,compsoc]{IEEEtran}

\usepackage{tabularx}
\usepackage{xcolor}
\usepackage{subfigure}
\usepackage{graphicx}
\usepackage{amsmath,amssymb} 
\usepackage{url} 
\usepackage{comment}
\usepackage[algoruled,medskip,dontprintsemicolon,linesnumbered,Algorithm]{algorithm}
\usepackage[]{algpseudocode}
\usepackage{psfrag}

\newcommand{\Fig}[1]{Fig.~\ref{fig:#1}}
\newcommand{\Sec}[1]{Sec.~\ref{sec:#1}}
\newcommand{\Tab}[1]{Tab.~\ref{tab:#1}}
\newcommand{\Eq}[1]{(\ref{eq:#1})}
\newcommand{\Alg}[1]{Alg.~\ref{alg:#1}}
\newcommand{\Line}[1]{Line~\ref{line:#1}}
\newcommand{\Nc}{\mathcal{N}}
\newcommand{\Kc}{\mathcal{K}}
\newcommand{\Tc}{\mathcal{T}}
\newcommand{\Bc}{\mathcal{B}}
\newcommand{\Gc}{\mathcal{G}}
\newcommand{\Dc}{\mathcal{D}}
\newcommand{\Fc}{\mathcal{F}}
\newcommand{\Ec}{\mathcal{E}}
\newcommand{\Pc}{\mathcal{P}}

\begin{document}

\title{
From Megabits to CPU~Ticks:\\
Enriching a Demand Trace in the Age of MEC
}

\author{Francesco~Malandrino,~\IEEEmembership{Member,~IEEE,}
Carla-Fabiana~Chiasserini,~\IEEEmembership{Fellow,~IEEE,}
Giuseppe~Avino, Marco~Malinverno,
Scott~Kirkpatrick,~\IEEEmembership{Life~Fellow,~IEEE}
\IEEEcompsocitemizethanks{\IEEEcompsocthanksitem F. Malandrino, C. F. Chiasserini, G. Avino and M. Malinverno are with Politecnico di Torino, Italy. C. F. Chiasserini is also with CNIT and a research associate with CNR-IEIIT, Torino, Italy. S. Kirkpatrick is with the Hebrew University of Jerusalem, Israel.
}
\thanks{This work is supported by the European Commission through the H2020 projects 5G-TRANSFORMER (Project ID 761536) and 5G-EVE (Project ID 815074).}}
\maketitle

\begin{abstract}
All the content consumed by mobile users, be it a web page or a live stream, undergoes some processing along the way; as an example, web pages and videos are transcoded to fit each device's screen. The recent multi-access edge computing (MEC) paradigm envisions performing such processing {\em within} the cellular network, as opposed to resorting to a cloud server on the Internet. Designing a MEC network, i.e., placing and dimensioning the computational facilities therein, requires information on how much computational power is required to produce the contents needed by the users. However, real-world demand traces only contain information on how much data is downloaded. In this paper, we demonstrate how to {\em enrich} demand traces with information about the computational power needed to process the different types of content, and we show the substantial benefit that can be obtained from using such enriched traces for the design of MEC-based networks.
\end{abstract}

\section{Introduction}
\label{sec:intro}

Dynamic web pages, targeted advertisement, user-generated content have all increased the {\em computation} required to assemble the pages displayed by web browsers. Mobile services, and the mobile devices consuming them, have made this trend even more evident, and virtually all content mobile users see is the result of multiple steps of real-time, on-the-fly processing. Prominent examples include videos, that are transcoded to match the screen size and resolution of the device playing them, and social networks, which decide what to show to their users based on their identity and location.

{\em Where} such processing ought to be performed is a fundamental question. The traditional approach was to use ad hoc servers placed within each content provider's network; cloud computing, with shared virtual servers placed in the Internet, provides a more efficient but substantially equivalent alternative~\cite{5gcloud}. The recent multi-access edge computing (MEC) paradigm takes a different approach: it envisions moving  services, i.e., the servers providing them, as close to mobile users as possible. Servers will not be localized in remote datacenters, but at different entities within the mobile core network itself, from core switches to base stations~\cite{noi-pof,noi-workshop}.
Similarly to cloud scenarios, there is a measure of cooperation between content providers and network operators. The servers deployed by the network operators will run services developed by the content providers, usually coming as a set of {\em virtual network functions} (VNFs) running on virtual machines or containers.

Designing a MEC network requires making decisions on where, within the network edge, servers shall be placed, and how to dimension them. Both questions require a deep knowledge of the data demand the network will have to serve: more exactly, we need to know (i) how much data the users will require; (ii) the type of such data, e.g., the mobile apps requesting it, and (iii) how much computational power will be needed to produce these data. The first two items have been widely studied: using operator-provided~\cite{orange-d4d,Malandrino-TCS} and crowd-sourced~\cite{noi-pof,noi-workshop} traces, it is possible for researchers to obtain a fairly good picture of the data demand of mobile networks, including its time and space evolution, as well as the services contributing to it.

Estimating the {\em processing power} needed to serve a given network demand, on the other hand, is much more challenging. The processing required to generate one gigabit per second of video data is not the same as the processing needed for the same quantity of gaming updates or maps. Additionally, it makes a significant difference whether the data is consumed by a small number of users enjoying a high bandwidth, or through a larger number of lower-rate connections. {\em None} of the currently available real-world traces contain all the information needed to distinguish these cases.

We fill this gap by developing and demonstrating a methodology to {\em enrich} existing demand traces with computational power information, translating (so to say) megabits of downloaded data into CPU ticks consumed at the servers. Specifically, in this paper we (i) consider a large-scale, crowd-sourced cellular demand trace,
already used in big data applications~\cite{scott}; (ii) perform an extensive set of experiments, linking the quantity of downloaded data in different conditions (type of service, number of users,...) to the amount of CPU power required at the servers; (iii) assess the impact of  enriching our traces on the resulting planning of the MEC network.
Importantly, we make {\em publicly available} realistic user demand traces, as well as our experimental dataset linking user demand to computational burden~\cite{html}.

We stress that, although MEC was our original motivation and we use it as a test case, the problem we address is more fundamental and consistent with the high-level goal of big data research, that is, to turn data into {\em information}. In our case, data come from real-world measurements, and the information we seek has to do with next-generation networks based on MEC.
In spite of  all their (perceived) abundance and the (apparent) easiness with which they are processed, real-world measurements
can only deal with present-day systems and, thus, they are not naturally well-suited to study future ones. In this paper, we demonstrate how this limitation can be overcome with the help of additional data, real-world experiments, and domain knowledge.

The remainder of the paper is organized as follows. We begin by reviewing related work in \Sec{relwork}. Then we present the real-world demand traces we use as a starting point in \Sec{traces}, and describe the experiments we perform to enrich them in \Sec{enrich}. \Sec{design} introduces and discusses our MEC network design strategy. After presenting our numerical results in \Sec{results}, we conclude the paper in \Sec{conclusion}.

\section{Related work}
\label{sec:relwork}

Our study is connected to three main categories of prior work: papers presenting real-world mobile traces and datasets; studies addressing MEC in general and MEC-based 5G networks specifically; works doing the latter using the first.

Many real-world traces come from volunteers, such as the MIT Reality Project~\cite{mit-reality} and the Nokia Mobile Challenge. These traces include a great deal of valuable information; their main shortcoming is the limited number of participants (in the case of the Nokia Mobile Challenge, around two hundred). This scale is adequate to study, for example, user mobility or encounter patterns, but studying a whole cellular network requires information about many more users.
Mobile operators are typically reluctant to release demand and deployment information to the scientific community. An exception is represented by the Data For Development dataset by Orange~\cite{orange-d4d}, including mobility information for 50,000 users in Ivory Coast, as well as CDR (call-detail record) information for phone calls and SMS messages. However, the Orange trace only includes voice calls and SMS, and is severely restricted by heavy anonymization -- each ID encountered gets a new coded identity for each ``ego site'' to which they are a neighbor. In other cases~\cite{Malandrino-TCS}, mobile operators have released demand or deployment information to individual research teams under non-disclosure agreements; however, these traces typically include only one operator and/or only one city.

MEC has been recently introduced~\cite{etsi-wp} as a way to move ``the cloud'', i.e., the servers processing mobile traffic, closer to the end users, thus reducing latency and traffic load across the network infrastructure. Network Function Virtualization (NFV) is widely regarded to as an enabling technology for MEC (see, e.g., \cite{etsi-wp}). Recent works have studied the radio techniques needed to enable MEC~\cite{mec-radio}, its relationship to the Internet-of-things~\cite{mec-iot} and context-aware, next-generation networks~\cite{mec-5g}. Closer to our scenario, the authors of~\cite{moving} study how caches and servers should be placed in the network as its load changes over time.
Regarding MEC and caching, a prominent application is mobile video streaming. As an example,~\cite{multiop-caching,cdn1} account for layered video coding techniques, and address the problem of placing the right layers at the right cache -- with~\cite{multiop-caching} also accounting for cooperation between operators. Other works~\cite{proactive-caching,wons} aim at {\em foreseeing} the content demand, in order to proactively populate caches or serve users.

5G will significantly exploit the MEC paradigm, and a substantial body of research is devoted to the problem of placing VNFs across the network providers' servers. As an example,\cite{AHirwe16,TKuo16,ABaumgartner15, FJemaa16,BAddis15} tackle the problems of VNF placement and routing from a network-centric viewpoint, i.e., they aim at minimizing the load of network resources. In particular,~\cite{AHirwe16} seeks to balance the load on links and servers, while~\cite{TKuo16} studies how to optimize routing to minimize network utilization. The above approaches formulate mixed-integer linear programming (MILP) problems and propose heuristic strategies to solve them. \cite{ABaumgartner15}, \cite{FJemaa16} and~\cite{BAddis15} formulate ILP problems, respectively aiming at minimizing the cost of used links and network nodes, minimizing resource utilization subject to QoS requirements, and minimizing bitrate variations through the VNF graph.

Not many works however exist that combine real-world traces and multi-access edge computing. Among the most recent ones, \cite{noi-pof}~studies the price (in terms of additional infrastructure) of deploying caches within the cellular core network. Compared to our work, \cite{noi-pof} only focuses on caching and vehicular traffic,  and it only considers the dataset for the city of Los Angeles.

Our earlier work~\cite{noi-workshop} sets in the same scenario as this paper: the traces introduced in \Sec{traces} were used to study the trade-offs between network latency and server utilization in MEC network design. The results presented in~\cite{noi-workshop} use the quantity of data downloaded by users as a proxy metric for computational capabilities required at the servers, and their limited applicability represents one of the main motivations behind this paper.

\section{Input data}
\label{sec:traces}

WeFi, now acquired by TruConnect Technologies, is an Android application providing location-specific information on the speed, security, and reliability of nearby Wi-Fi networks. At the same time, it collected data on the activity of its users, including location, mobile phone activity, and available connectivity options. For our study we use three datasets, coming from the U.S. cities of Atlanta, Los Angeles, and San Francisco, characterized by the features summarized in \Tab{datasets}.

\begin{table}[t]
\caption{
The WeFi datasets
\label{tab:datasets}
} 
\begin{tabularx}{1\columnwidth}{|c|X|X|X|}
\hline
& Atlanta & Los Angeles & San Francisco \\
\hline\hline
Time of collection & Oct. 2015 & Oct. 2015 & Mar. 2015\\
\hline
Covered area [km$^2$] & $55\times 66$ & $46\times 73$ & $14\times 11$\\
\hline
Total traffic [TB] & 9.34 & 35.61 & 9.18\\
\hline
Number of records & 13 million & 81 million & 60 million\\
\hline
Unique users & 9,203 & 64,386 & 14,018\\
\hline
Unique cells & 12,615 & 36,09 & 14,728\\
\hline
\end{tabularx}
\vspace{-5mm}
\end{table}

Datasets are organized in {\em records}, each containing: time and GPS location; anonymized user identifier; current operator and cell identifier; active application on the smartphone and amount of data it downloads. New records are generated every time any of the above changes (e.g., the user moves or switches apps), or a one-hour period elapses.

The WeFi datasets represent a real-world, {\em live} snapshot of both mobile networks and their users, and have three features that make them especially relevant to our study. First, they contain information on several cities, different for traffic demand profile and network deployment. Furthermore, they include multiple mobile operators, with different deployment strategies, e.g., usage of micro- and macro-cells. Finally, they allow us to know the individual application generating each traffic flow.

The WeFi data has been already used in the field of {\em big data}, e.g., in the preliminary analysis~\cite{scott} and follow-up works. The purpose therein was to identify patterns of living, commuting, fast food consumption, work activities and recreation for tens of thousands of people. That case study confirms that the content and quality of the WeFi datasets are sufficient not only for networking studies, but also for social observations and government actions.

{\bf Dataset available.} While we cannot share the dataset we use, we produced a synthetic trace with the same time and space characteristic of its original counterpart, available for download from~\cite{html}. For more details on the generation of such synthetic traces, the interested reader is referred to~\cite{noi-tmc17}.

\section{Enriching the mobile data traces}
\label{sec:enrich}

As mentioned in \Sec{intro}, our high-level goal is to use the {\em demand} information available in the datasets to understand the {\em deployment} we need, i.e., how much computational capacity shall be placed within the network, and where. To this end, we perform three main steps:
\begin{enumerate}
\item we design and perform a set of experiments measuring the computational load associated with different types of traffic, as detailed in \Sec{sub-exp};
\item we use the experimental data to train a {\em model} connecting the two quantities, as described in \Sec{sub-learn};
\item we exploit the model to estimate the computational capacity needed to serve the traffic we observe in our dataset, as discussed in \Sec{sub-enrich}.
\end{enumerate}

\subsection{Experimental setup}
\label{sec:sub-exp}

We focus on three different, highly representative types of traffic, namely, video streaming, gaming, and maps. For reproducibility purposes, we restrict ourselves to open-source programs, namely:
\begin{itemize}
\item for video, the FFserver server and the VLC client;
\item for gaming, the Minecraft server and the corresponding Minecraft Pocket mobile client;
\item for maps, the OpenMapTiles server~\cite{tiles-server}, based on data from OpenStreetMap~\cite{osm}.
\end{itemize}

For our experiments, we use a testbed, composed of two computers -- a client and a server -- connected via a Wi-Fi link. Both the client and the server are commodity, off-the-shelf computers equipped with Intel Core~i7-7700T processors and running Ubuntu Linux~16.10. On the client machine, we run the Genymotion Android emulator, which in turn emulates a varying number of mobile clients. The server machine hosts one instance of the server application, containerized within Docker.

In streaming tests, we use three videos with different duration, resolution, and file size; for gaming, we employ three different moving patterns throughout the Minecraft world; for maps, we generate requests for randomly-chosen $20\times 20$\,km$^2$-areas. In both cases, we vary the number of clients between~1 and~8, and use the FRep program to drive the emulator, thus guaranteeing uniform, reproducible patterns of UI actions, e.g., taps and swipes.

In all experiments, we measure two quantities: the amount of data generated by the server program, and the computational capacity it consumes. The former simply corresponds to the traffic outgoing from the \path{docker0} virtual interface created by Docker; the latter is obtained by polling the \path{/proc/<PID>/stat} file relating to the server process.
The FFserver and Minecraft servers are single-threaded, so there is only one process to account for. The OpenMapTiles server, on the other hand, includes several processes, including a database and a web server; in that case, we present the aggregate CPU consumption figures.

\subsection{Results and model}
\label{sec:sub-learn}

\begin{figure*}
\centering
\subfigure[\label{fig:scatter-video}]{
\includegraphics[width=.3\textwidth]{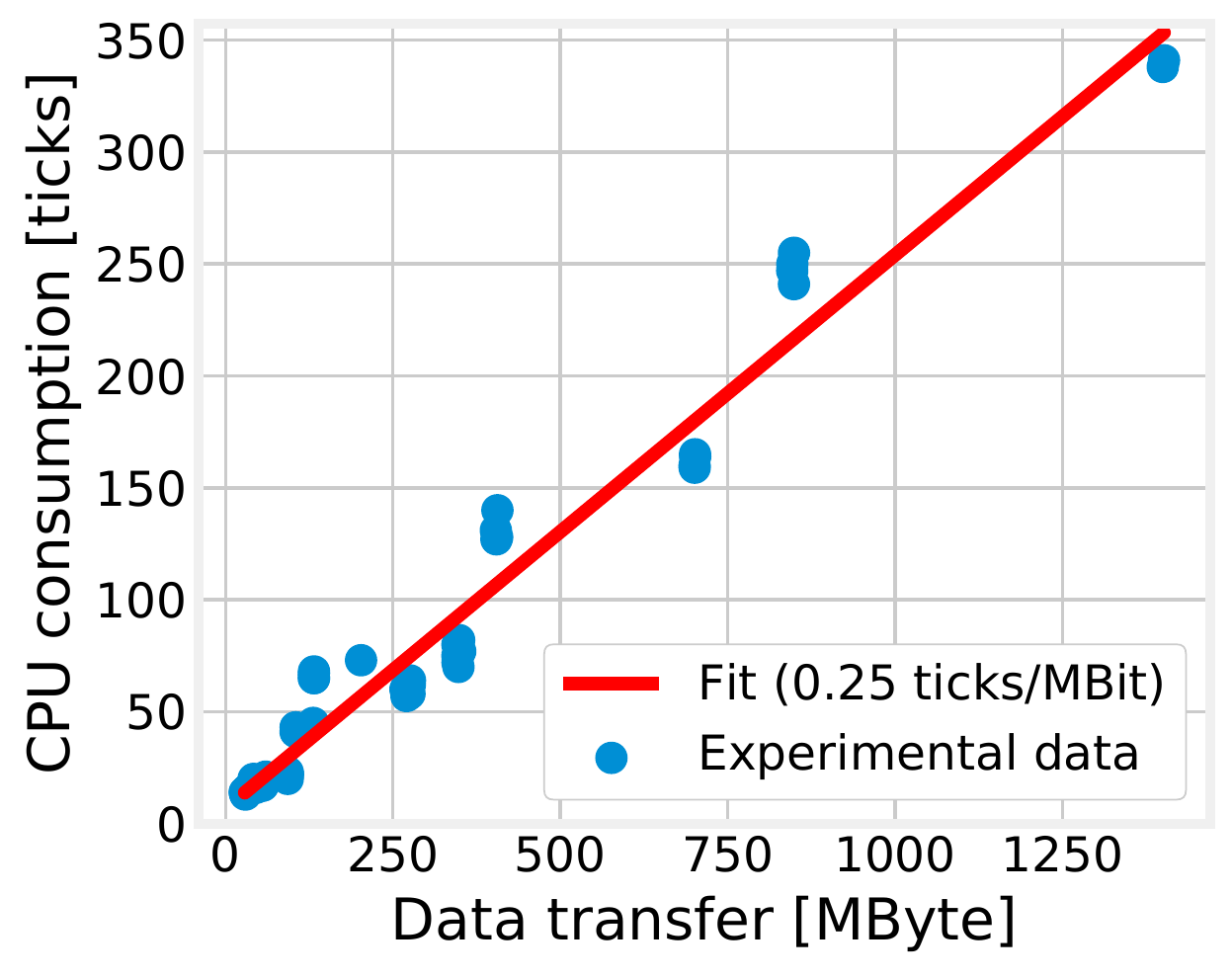}
} 
\subfigure[\label{fig:scatter-gaming}]{
\includegraphics[width=.3\textwidth]{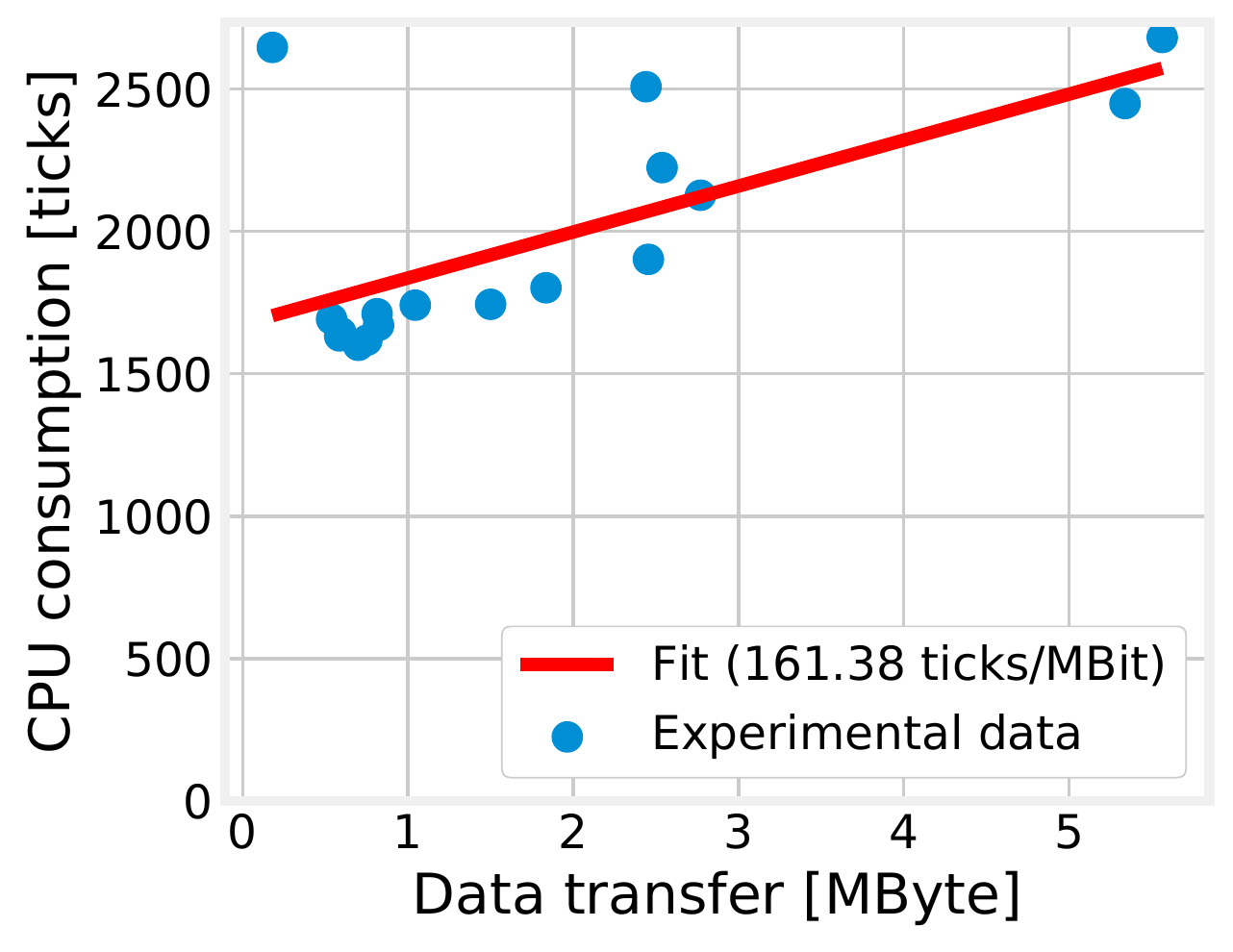}
} 
\subfigure[\label{fig:scatter-maps}]{
\includegraphics[width=.3\textwidth]{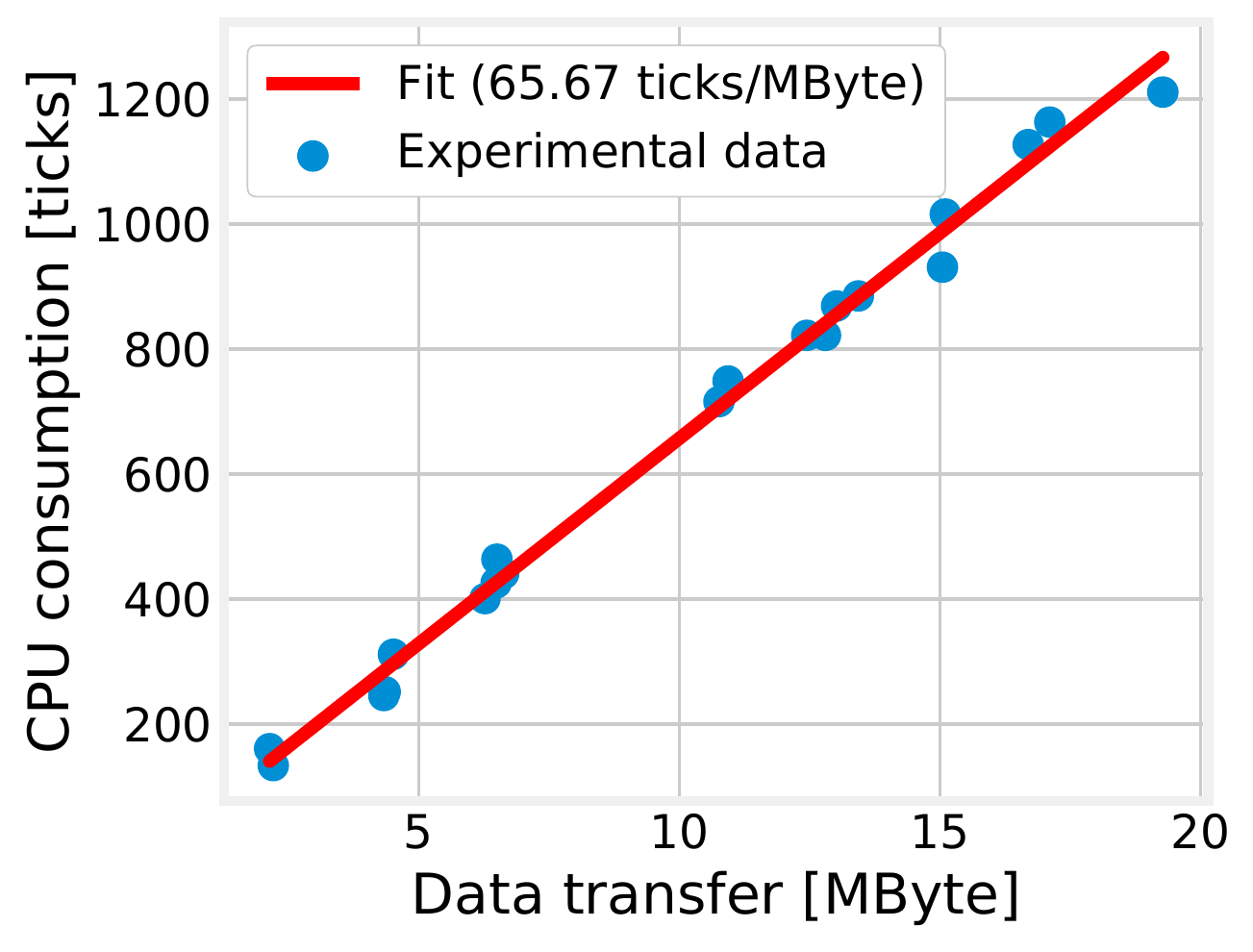}
} 
\caption{
Data (blue dots) and linear fit (red line) for the video (a), gaming (b) and maps (c) applications. Fitting error (RMSE) are 4\%, 6\% and 2\% respectively.
\label{fig:scatter}
} 
\end{figure*}

\Fig{scatter-video}, \Fig{scatter-gaming} and \Fig{scatter-maps} summarize our experimental results. Each blue dot therein corresponds to an experiment, e.g., a different combination of video file and number of clients; its position in the plot is determined by how much data was served in that experiment, and the corresponding load on the server. Computational load values are expressed in {\em CPU ticks}, the minimum unit of CPU scheduling in the operating system. In Linux, each tick represents 10~ms; as an example, saying that a process takes 200~ticks to complete means that the CPU was assigned to that process for 2~s \footnote{The real execution time might be longer, e.g., if the process is preempted by a higher-priority one during its execution.}.

The first thing to notice is the scale of the two plots. As one might expect, serving video implies transferring massively more data than hosting a game server, and therefore video is rightfully regarded as one of the applications consuming most of the bandwidth offered by networks.
Looking at the y-axis, on the other hand, a different aspect emerges: the amount of CPU resources associated with gaming and, to a lesser extent, maps, dwarfs the one required by video.

This disproportion makes intuitive sense: while streaming a video requires little more than reading a file from disk and feeding it into a network stream, game and map servers have to perform complex operations such as keeping track of the game status or rendering the maps. However, its scale is somehow unexpected, and serves as a clear reminder that designing and deploying the {\em computation} part of a network, i.e., where and how to place its computational power, using only the traffic demand as a guideline is likely to result in poor performance and low efficiency.

The vastly different relationships between served traffic and CPU load are also evident from the linear fit we obtain from the two sets of experiments, represented by red lines in \Fig{scatter}. The fitted relationships are as follows:
\begin{equation}
\label{eq:model}
\text{CPU}_{[\text{ticks}]}=\begin{cases}
0.25\times\text{traffic}_{[\text{Mbit}]}+6.76 & \text{for video,} \\
161.38\times\text{traffic}_{[\text{Mbit}]}+1675.03 & \text{for gaming,} \\
67.44\times\text{traffic}_{[\text{Mbit}]}-7.53 & \text{for maps.} \\
\end{cases}
\end{equation}
Looking at \Fig{scatter}, it also important to observe how the fit is much better for maps traffic than for gaming: the root mean square error (RMSE) for maps is 2\% while it is equal to 6\% for gaming.  The reason is that the quantity of CPU required by gaming applications significantly depends on such factors as the actions performed by the players -- a fact that leads to a larger variability.

The fact that all fitted relationships are linear is not especially relevant {\em per se}, nor particularly surprising; indeed, it makes intuitive sense that the amount of work needed to produce a certain type of data grows, more or less linearly, with the quantity of data to produce. What matters the most is the {\em slope} of the fitted lines, which changes by several orders of magnitude from one traffic category to another.

In the following, we will indicate with~$\tau_k$ the number of CPU ticks needed to process one  megabyte of traffic of category~$k$; for instance, $\tau_\text{video}=0.25$~CPU ticks/Mbyte (i.e., the slope of the red line in \Fig{scatter-video}), $\tau_\text{gaming}=161.38$~ticks/Mbyte (i.e., the slope of the red line in \Fig{scatter-gaming}), and $\tau_\text{maps}=67.44$~ticks/Mbyte (i.e., the slope of the red line in \Fig{scatter-maps}).

\subsection{Enriching the dataset}
\label{sec:sub-enrich}

\begin{figure}
\centering
\includegraphics[width=.23\textwidth]{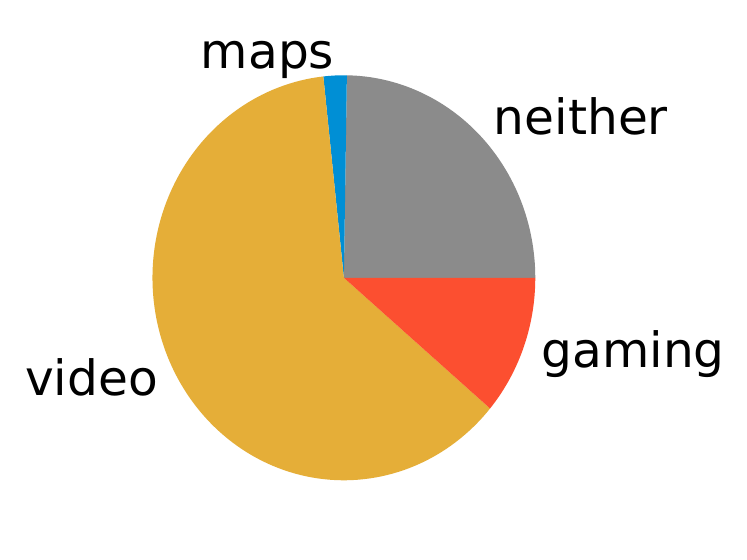}
\includegraphics[width=.23\textwidth]{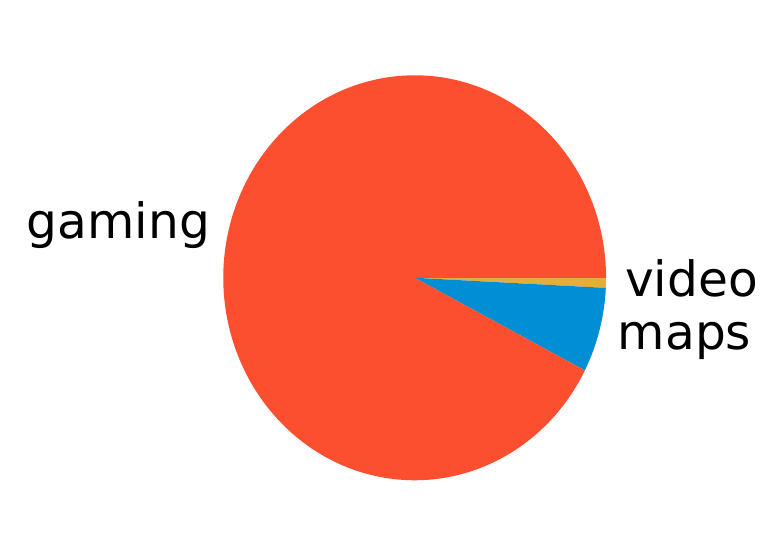}
\caption{
Categories traffic demand belongs to (left); amount of computational power required to serve video-, and gaming- and maps-like traffic (right).
\label{fig:pies}
\vspace*{-3mm}
} 
\end{figure}

As a preliminary step, we need to decide, for each app we observe in the dataset, whether its traffic can be considered video-like, gaming-like, or map-like (or none of them). We perform such an assignment as follows:
\begin{itemize}
\item traffic coming from YouTube, Netflix, TimeWarner, ShowBox, Twitch, DirectTV, FoxSports, FoxNews, is tagged as video-like;
\item traffic coming from Minecraft, World of Warcraft, Riptide, Grand Theft Auto, Rollercoaster Tycoon, This War of Mine, Titan Quest, Unkilled, is tagged as game-like;
\item traffic coming from Google Maps and Waze is tagged as map-like.
\end{itemize}
As summarized in \Fig{pies}(left), about 66\% of the traffic we observe in our dataset can be classified as video-like, about 15\% of it as game-like, while the amount of map-like traffic is much smaller. Much of the traffic that cannot be tagged comes from social networking applications, whose behavior cannot be easily studied due to the lack of open-source social network servers.

Once the assignment is made, we can use the relationship in \Eq{model}, learned in \Sec{sub-learn}, to add a new column to our dataset. The column, called \path{cpu_ticks}, expresses how much computational power is required to generate the traffic reported in each line of the dataset.
Notice that relationships other than linear can be accounted for at no additional complexity.

\Fig{pies} further highlights the gap between the quantity of traffic generated by different applications and the corresponding CPU load.
Gaming and maps represent a small fraction of the total traffic (\Fig{pies}(left)); however, by applying \Eq{model}, we obtain the results shown in \Fig{pies}(right): gaming applications consume the vast majority, over 90\%, of all computational resources. This further highlights the importance of accounting for the computational load of different types of traffic in network design.

\subsection{Discussion}
\label{sec:sub-discuss}

Our trace enrichment strategy has some limitations: it does not account for all the traffic types, and might not perfectly model the behavior of all servers.

The first issue is evident from \Fig{pies}(left): around 25\% of all traffic cannot be classified as either video-like,
gaming-like, or maps-like. On the other hand, our analysis is able to account for over 70\% of all present-day traffic, and video, gaming and maps are the applications that are expected to grow the most in the near future.

The second issue has to do with the fact that we model the behavior of YouTube and Netflix using FFserver, of World of Warcraft through Minecraft, or of Google Maps using OpenStreetMap. On the one hand, this might sound a bit bold. On the other hand, it is true that the problems faced by different applications of the same type -- and the solutions thereto -- tend to overlap. This explains, as an example, the existence of gaming {\em engines} such as Unity or Microsoft XNA, providing the developers of a heterogeneous set of games with homogeneous solutions to a small set of common problems.

In summary, our analysis
provides valuable guidelines highlighting  the existence of a mismatch between the quantity of traffic and the corresponding CPU load, as well as its magnitude.
Additionally, the methodology we employ is general, and works unmodified in cases where additional traffic types and/or applications are taken into account, or if larger-scale experiments can be performed.

{\bf Dataset available.} The results of our experiments are available for download from~\cite{html}.

\section{MEC design}
\label{sec:design}

We now show how the enhanced dataset we created can be used to devise a MEC design strategy.
Our input data are represented by:
\begin{itemize}
\item a set of base stations;
\item the expected/predicted traffic at each of them and its characteristics;
\item the network connectivity.
\end{itemize}
Given the above, we have to dimension the network computational capabilities, i.e., to decide (i) where to place the MEC servers, and (ii) the capacity they should have.

We solve this problem accounting also for the specific applications we deal with, thus developing an {\em application-aware deployment},
accommodating the requirements of each application. Indeed, such requirements can be substantially different for different {\em categories}
of applications. As an example, we might be willing to serve such (comparatively) delay-tolerant content as video-like and map-like traffic through a server located in the core network, while real-time services like mobile gaming will require their servers to be much closer to the end users, possibly at individual base stations.
It is important to stress that, in pure NFV/MEC fashion, we allow the same servers to serve multiple applications concurrently, provided that the server capacity is not exceeded.

At last, note that our input data is represented by the traffic that is {\em actually} served by the base stations. Consistently with MEC design best practices~\cite{mec-survey,mec-ton}, we do not need to explicitly account for access-network issues such as congestion and interference. Indeed, the coverage quality experienced by users influences the amount of data they are able to upload/download, and this aspect is captured by our system model and MEC design.

\subsection{Network topology}

The identity and positions of base stations~$b\in\Bc$, as well as their demand~$\delta(b,k,t)$ are readily available from the trace we describe in \Sec{traces}. However, we have no information about the topology of the backhaul cellular network. Indeed, mobile operators are extremely reluctant to disclose this information, and virtually all works in the literature resort to synthetic topologies based on current best practices. Following~\cite{softcell}, we assume a {\em fat-tree} topology, where:
\begin{itemize}
\item base stations are grouped into {\em rings} of ten;
\item every ten rings, there is an aggregation-level {\em pod};
\item every ten pods, there is a core-level switch.
\end{itemize}
The topology has a fan-out of~$2$, i.e., we connect every ring to the two closest pods, and every pod to the two closest core switches, while switches themselves are connected in a full mesh. The network topology we generate can be represented as a graph~$\Gc=(\Nc,\Ec)$.

We also generate a DAG~$\Dc=(\Nc,\Fc)$, to keep track of which nodes can serve each base station. In particular, we generate a (directed) edge~$(n_1,n_2)\in\Fc$ between nodes such that:
\begin{itemize}
\item $n_1$ and~$n_2$ are connected, i.e.,~$(n_1,n_2)\in\Ec$, and
\item $n_1$ belongs to a network level immediately higher to the one of~$n_2$, e.g., $n_1$ is a ring and $n_2$~is a base station.
\end{itemize}
We will say that node~$n_1$ is a {\em parent} of~$n_2$ if~$(n_1,n_2)\in\Fc$. Base stations have no children, i.e., they are leaves in the DAG, while core switches have no parents, i.e., they are roots.

{\bf Notation.}
We denote by~$n\in\Nc$ all the nodes of the cellular network, from base stations to core-level switches, and by~$b\in\Bc\subset\Nc$ the base stations among them. The physical distance between any two nodes~$n_1$ and~$n_2$ is denoted by~$d(n_1,n_2)$. Contents are specific to  {\em category}~$k\in\Kc$, e.g., video, gaming, or maps, and time is discretized into steps~$t\in\Tc$. We denote by $\delta(b,k,t)$ the demand from users covered by base station~$b$, for contents of category~$k$ and during step~$t$.

\subsection{MEC design}

Designing our MEC network means making two decisions, each corresponding to a binary variable:
\begin{itemize}
\item whether we should place a server at node~$n\in\Nc$, expressed through variable~$y(n)\in\{0,1\}$;
\item whether the traffic coming from base station~$b\in\Bc$ shall be served by the server placed at node~$n\in\Nc$, expressed through variable~$x(b,n)\in\{0,1\}$.
\end{itemize}
Note that none of the decision variables depends on the time step~$t\in\Tc$; this reflects the fact that deployment decisions are made periodically, with a time period much longer than a single time step, accounting for the evolution of data demand over the previous period.
Needless to say, we cannot serve anything on servers that do not exist, i.e., it must be:
\begin{equation}
\nonumber
x(b,n)\leq y(n),\quad\forall b\in\Bc,n\in\Nc.
\end{equation}

The objective of the problem can be stated as deploying the smallest possible number of servers subject to delay constraints, i.e.,
\begin{equation}
\nonumber
\min_{x,y}\sum_{n\in\Nc}y(n).
\end{equation}

\begin{algorithm}[t]
\caption{Greedy MEC design
\label{alg:design}
} 
\begin{algorithmic}[1]
\Require{$k,L_{\max},\delta(b,k,t),\Bc,\Nc$} \label{line:input}
\ForAll{$b\in\Bc$} \label{line:init}
\If{$\max_{t\in\Tc}\delta(b,k,t)>0$}
\State{$y(b)\gets 1$}
\State{$x(b,b)\gets 1$}
\EndIf
\EndFor
\While{$\texttt{latency}\leq L_{\max}$} \label{line:while}
\State{$\Pc\gets\{(n_1,n_2)\in\Nc\colon y(n_1)>0 \wedge y(n_2)>0 \wedge (n_1,n_2)\in\Fc\}$} \label{line:pairs1}
\State{$\Pc\gets\Pc\cup\{(n_1,n_2)\in\Nc\colon y(n_1)>0 \wedge y(n_2)>0 \wedge \exists n_3\in\Nc\colon(n_3,n_1)\in\Fc\wedge(n_3,n_2)\in\Fc\}$} \label{line:pairs2}
\State{$n_1^\star,n_2^\star\gets\arg\max_\Pc \texttt{score}(n_1,n_2)$} \label{line:mindist}
\State{$\textbf{consolidate}(n_1^\star,n_2^\star)$} \label{line:call}
\EndWhile
\State\Return{$x(b,n),y(n)$}
\end{algorithmic}
\end{algorithm}

\begin{algorithm}[t]
\caption{The consolidation procedure
\label{alg:consolidate}
} 
\begin{algorithmic}[1]
\Require{$n_1,n_2,k$}
\If{$(n_1,n_2)\in\Fc$} \Comment{parent-children, like \Fig{consolidate}(top)}\label{line:consolidate1-begin}
\ForAll{$b\in\Bc\colon x(b,n_2)>0$}
\State{$x(b,n_2)\gets0$}
\State{$x(b,n_1)\gets1$}
\EndFor
\State{$y(n_2)\gets0$} \label{line:consolidate1-end}
\Else \Comment{siblings, like \Fig{consolidate}(bottom)} \label{line:consolidate2-begin}
\State{$n_3\gets n_3\in\Nc\colon (n_1,n_3)\in\Fc\wedge(n_2,n_3)\in\Fc$} \label{line:consolidate2-n3}
\ForAll{$b\in\Bc\colon x(b,n_1)>0$}
\State{$x(b,n_1)\gets0$}
\State{$x(b,n_3)\gets1$}
\EndFor
\ForAll{$b\in\Bc\colon x(b,n_2)>0$}
\State{$x(b,n_2)\gets0$}
\State{$x(b,n_3)\gets1$}
\EndFor
\State{$y(n_1)\gets0$}
\State{$y(n_2)\gets0$}
\State{$y(n_3)\gets1$} \label{line:consolidate2-end}
\EndIf
\end{algorithmic}
\end{algorithm}

Optimally setting the binary $x$- and $y$-variables subject to constraints on the need to serve all traffic requires solving an ILP problem\footnote{We skip the proof, based on a reduction from the SAT problem, in the interest of space.}, which is notoriously~\cite{milp} impractical even for modestly-sized problem instances. We therefore devise a greedy design procedure, able to efficiently make good-quality deployment decisions.

\subsubsection{Greedy design procedure}
\label{sec:algos}

Our greedy design procedure
is inspired to hierarchical clustering and summarized in \Alg{design}. It takes as input (\Line{input}) the content category~$k$ to consider, its maximum processing latency~$L_{\max}$, its demand over time~$\delta(b,k,t)$, and the sets~$\Bc$ and~$\Nc$ of, respectively,  base stations and network nodes. This implies that \Alg{design} makes {\em per-category} decisions, and is thus able to account for the fact that different categories of content, with different latency limits~$L_{\max}$, might require different deployment strategies. Notice that with {\em latency} we indicate the time spent within the core network, which is itself a component of the total, end-to-end service time.

The algorithm
starts (\Line{init}) from a solution where each base station
that ever serves at least one user requiring content category~$k$ (i.e., $\max_{t\in\Tc}\delta(b,k,t)>0$) has its own server, i.e., $y(b)=1$.
Then, as long as the latency resulting from our deployment does not exceed the threshold~$L_{\max}$ (\Line{while}), we select a pair of nodes~$(n_1,n_2)$ to {\em consolidate} together, reducing the number of servers at the price of potentially increasing the service latency.

\begin{figure}[t]
\psfrag{n2}[m][c]{$n_1$}
\psfrag{n1}[m][c]{$n_2$}
\psfrag{n3}[m][c]{$n_3$}
\centering
\includegraphics[width=.22\textwidth]{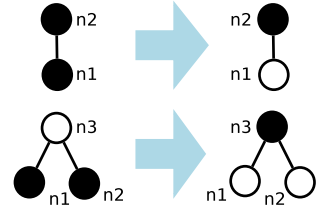}
\caption{
The consolidation procedure. Circles represent nodes in~$n\in\Nc$; black ones correspond to nodes that host a server, i.e.,~$y(n)=1$, white ones to nodes that do not, i.e.,~$y(n)=0$. If $n_1$ is $n_2$'s parent (top), the server at~$n_2$ is removed and all base stations it served, are served by $n_2$'s parent~$n_1$. If $n_1$ and $n_2$ are siblings (bottom), the servers at~$n_1$ and~$n_2$ are removed, and a new one is created at their parent~$n_3$.
\label{fig:consolidate}
\vspace{-3mm}
} 
\end{figure}

The consolidation procedure is depicted in \Fig{consolidate}. It involves two nodes~$n_1$ and~$n_2$ such that either $n_2$ is $n_1$'s parent, or $n_1$ and~$n_2$ are siblings, i.e., have a common parent~$n_3$. In both cases, the server(s) at the child(ren) are removed, and all base stations they used to serve are served by the parent. Every time we perform the consolidation procedure, the number of servers deployed in the topology decreases by one unit, at the cost of potentially increasing the latency.

In \Line{pairs1}--\Line{pairs2} of \Alg{design}, we construct a set~$\Pc$ of pairs of nodes eligible for consolidation, i.e., such that (i) they both have a server, and either (ii) $n_1$~is $n_2$'s parent (\Line{pairs1}), or (iii) both~$n_1$ and~$n_2$ have a common parent~$n_3$. In \Line{mindist}, we select the nodes~$n_1^\star$ and~$n_2^\star$ with
the highest score, i.e., the most suitable to consolidate,
and call the {\bf consolidate} procedure with those nodes as an argument.
As discussed in \Sec{dist}, different definitions of score can be considered, leading  to different deployment strategies.

\Alg{consolidate} details the consolidation procedure, and takes as input the nodes~$n_1$ and~$n_2$ to consolidate. If $n_2$ is~$n_1$'s parent, i.e., we are in the situation of \Fig{consolidate}(top), the server at~$n_2$ is removed and any base stations that were served by~$n_2$ are served by~$n_1$ (\Line{consolidate1-begin}--\Line{consolidate1-end} of \Alg{consolidate}). If~$n_1$ and~$n_2$ are siblings, i.e., we are in the situation of \Fig{consolidate}(bottom), then
we first identify the node~$n_3$ that is a parent to both~$n_1$ and~$n_2$ (\Line{consolidate2-n3}). Afterwards, servers at both~$n_1$ and~$n_2$ are removed, a new server at~$n_3$ is created, and all base stations that were served by~$n_1$ or~$n_2$ are served by~$n_3$ (\Line{consolidate2-begin}--\Line{consolidate2-end} of \Alg{consolidate}).

\subsubsection{Score definitions}
\label{sec:dist}

A key feature of our approach is that it can support multiple deployment {\em strategies} through the one {\em algorithm} \Alg{design}. We are able to do so by considering multiple definitions of the score associated with consolidating two network nodes, i.e., the implementation of function \path{score} called in \Line{mindist}.
Recall that we are still focusing on a single content category~$k$.

\noindent{\bf Location-based}
It is often desirable to consolidate nodes that are physically close to each other, as this typically translates into a shorter travel time between the users and the servers serving them. This corresponds to giving higher scores to pairs of nodes that are close to each other, i.e.,
\begin{equation}
\label{eq:dist-dist}
\texttt{score}(n_1,n_2)=-d(n_1,n_2).
\end{equation}

\noindent{\bf Load-based}
An alternative deployment strategy takes into account the load of each server, and tries to avoid consolidating nodes whose demands
have a similar time evolution, i.e., whose peak times tend to overlap. The rationale is that by doing so we can decrease the capacity requirements for the consolidated servers, which depend upon the {\em peak} of the combined load.
More formally, let us define a serve vector~$\vec{s}(n)$ for each node~$n$. $\vec{s}(n)$~vectors have~$|\Tc|$ elements, and each element~$s(n)_{t}$ represents the total demand for contents of category~$k$ by users at base stations that are served by node~$n$ during time step~$t$:
\begin{equation}
\nonumber
s(n)_t=\sum_{b\in\Bc}x(b,n)\delta(b,k,t).
\end{equation}
Given the $\vec{s}(n)$~values, we can define the score related to the $(n_1,n_2)$ pair as
\begin{multline}
\label{eq:dist-load}
\texttt{score}(n_1,n_2)=\max_{t\in\Tc} \tau_k s(n_1)_t+\max_{t\in\Tc} \tau_k s(n_2)_t+\\
-\max_{t\in\Tc}\tau_k\left (s(n_1)_t+s(n_2)_t\right).
\end{multline}
where the first two terms of the second member of \Eq{dist-load} represent the peak loads of nodes~$n_1$ and~$n_2$ before consolidation; the third term is the peak load of the combined server, after consolidation. A high-scoring consolidation operation will involve nodes with high peak loads (first two terms, with positive sign) that can be combined into a new, low-load server (third term, with negative sign).

Recall that the factor~$\tau_k$ in \Eq{dist-load} expresses how many units of computational power (e.g., CPU ticks) are needed to generate one unit of traffic (e.g., one megabyte) of category~$k$, as obtained in \Sec{sub-enrich}. Using non-enriched traces corresponds to assuming~$\tau_k=1,\,\forall k\in\Kc$.

\begin{figure}
\centering
\includegraphics[width=.24\textwidth]{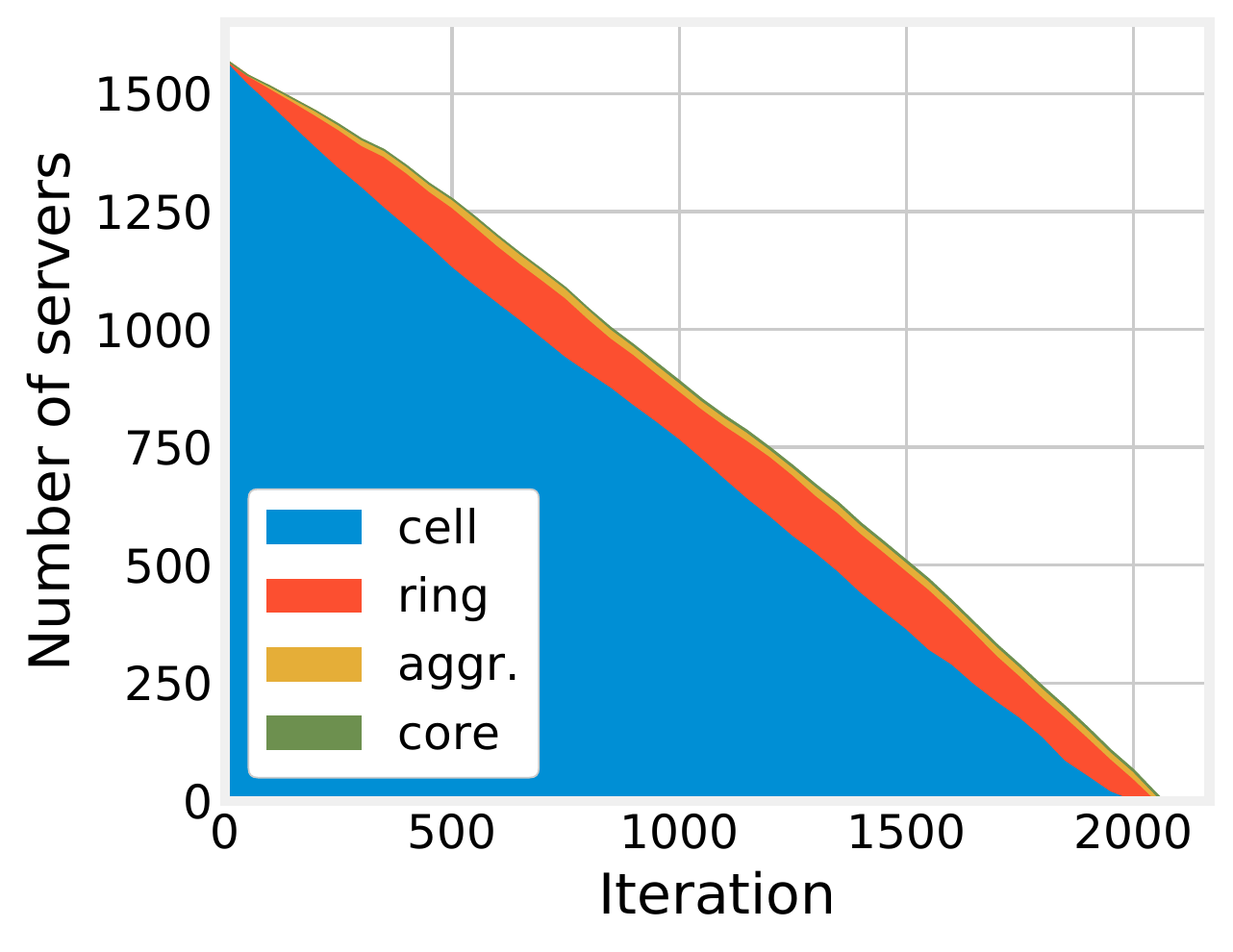}
\hspace{-4mm}
\includegraphics[width=.24\textwidth]{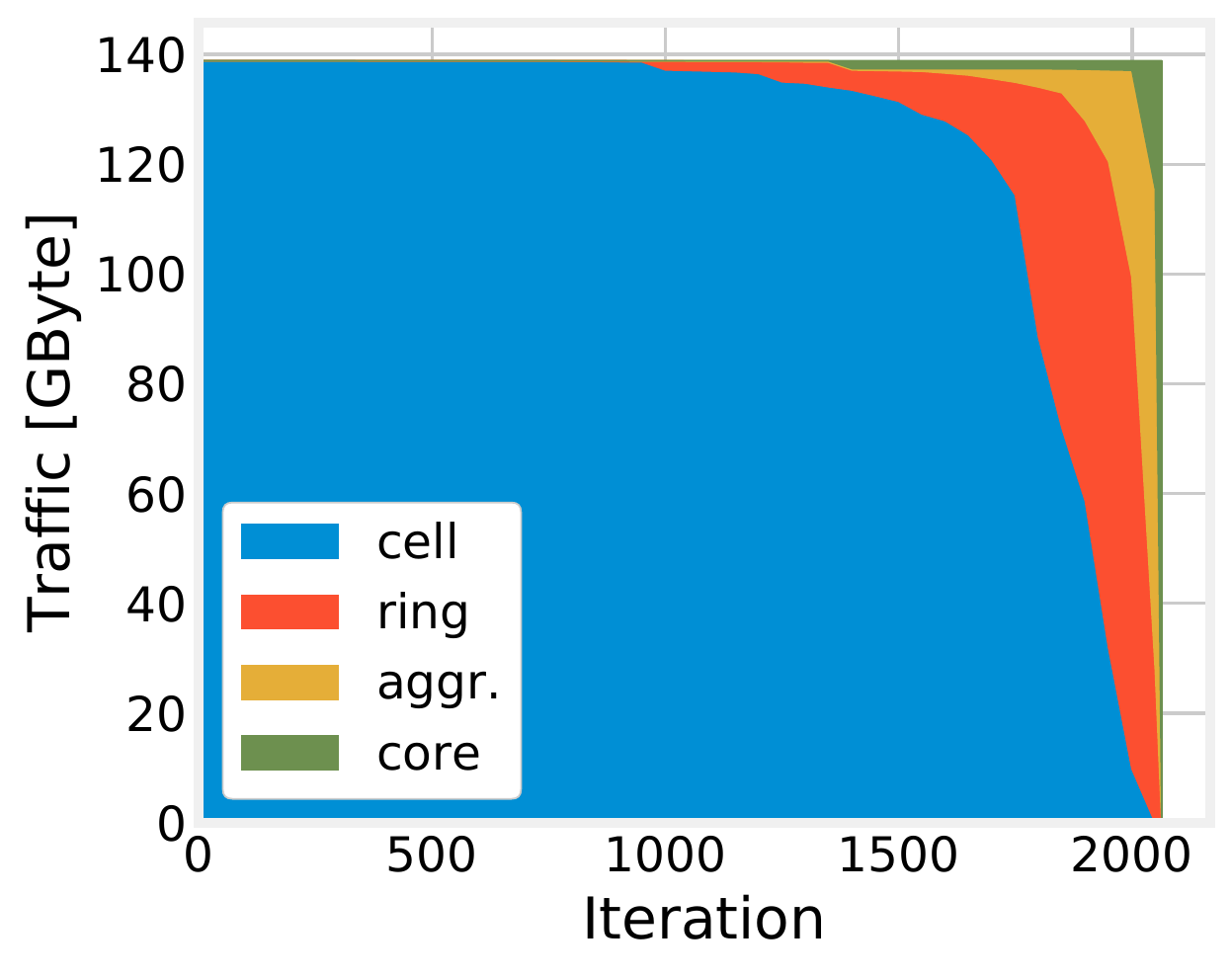}
\caption{\label{fig:fillnsrv-loc}Location-based scores: number of servers (left) and amount of traffic processed (right) at the different network levels, for each iteration of \Alg{design}.
} 
\end{figure}
\begin{figure}
\includegraphics[width=.24\textwidth]{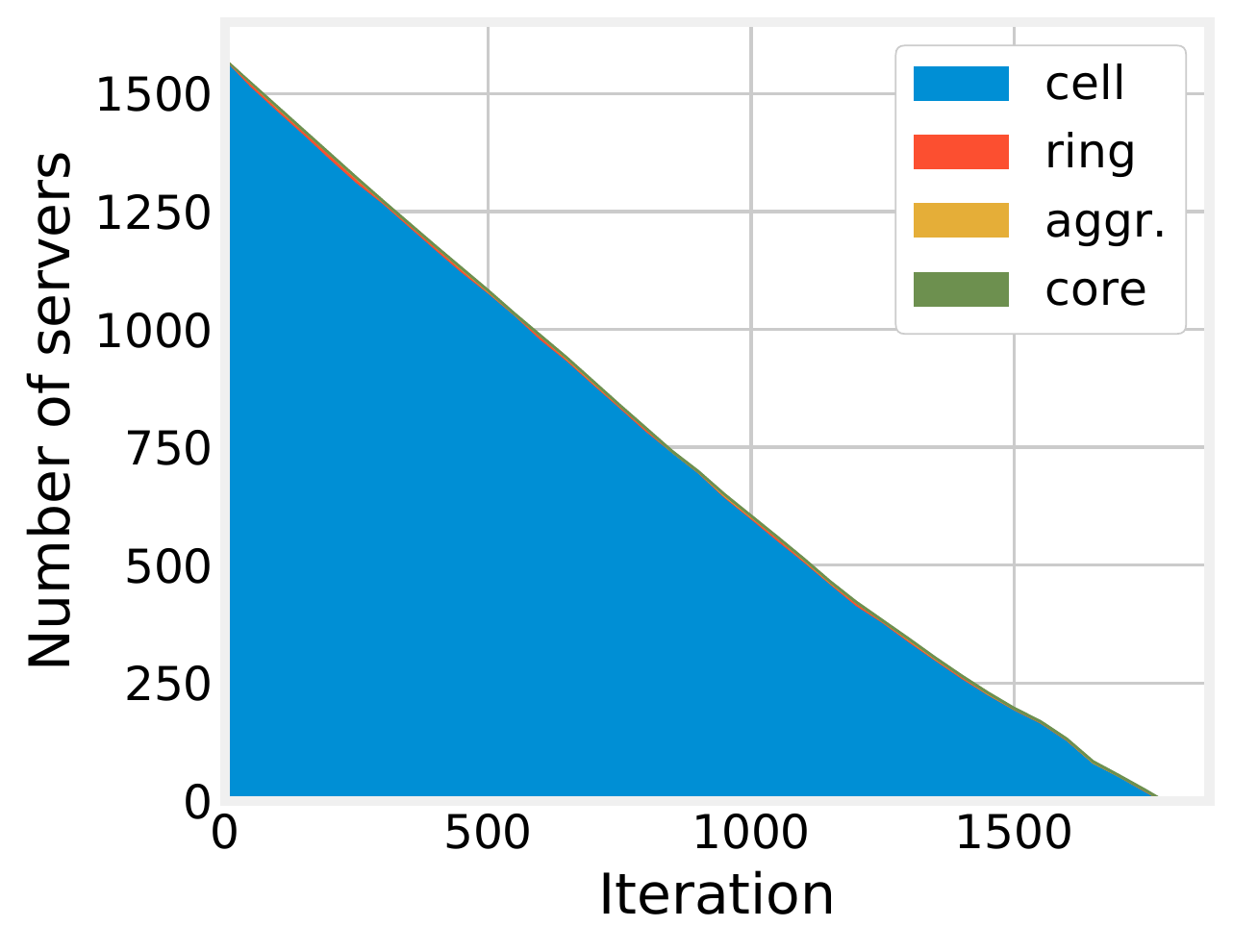}
\hspace{-4mm}
\includegraphics[width=.24\textwidth]{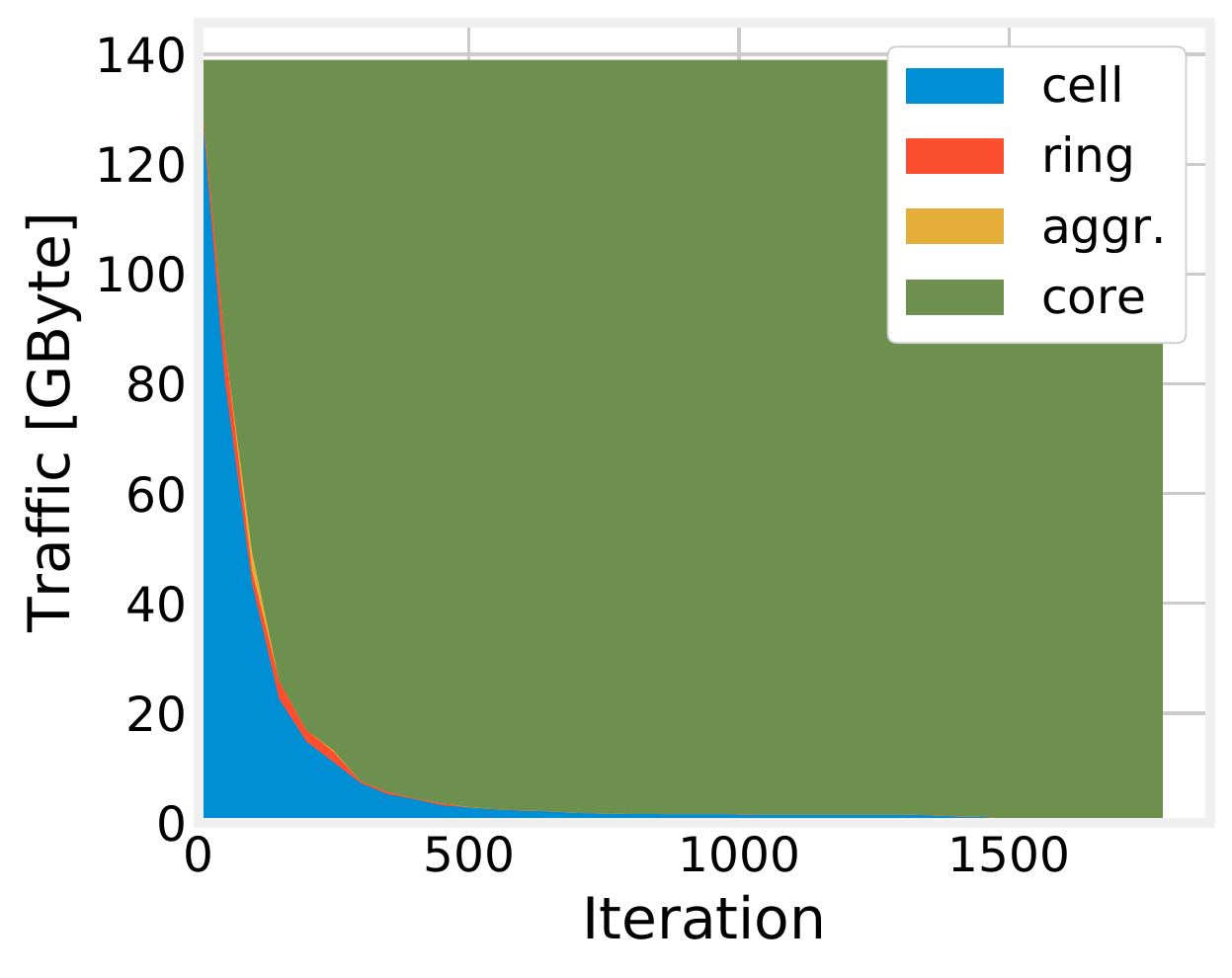}
\caption{\label{fig:fillnsrv-dem}Load-based scores: number of servers (left) and amount of traffic processed (right) at the different network levels, for each iteration of \Alg{design}.
} 
\end{figure}

\begin{figure*}
\centering
\subfigure[\label{fig:delay}]{
\includegraphics[width=.3\textwidth]{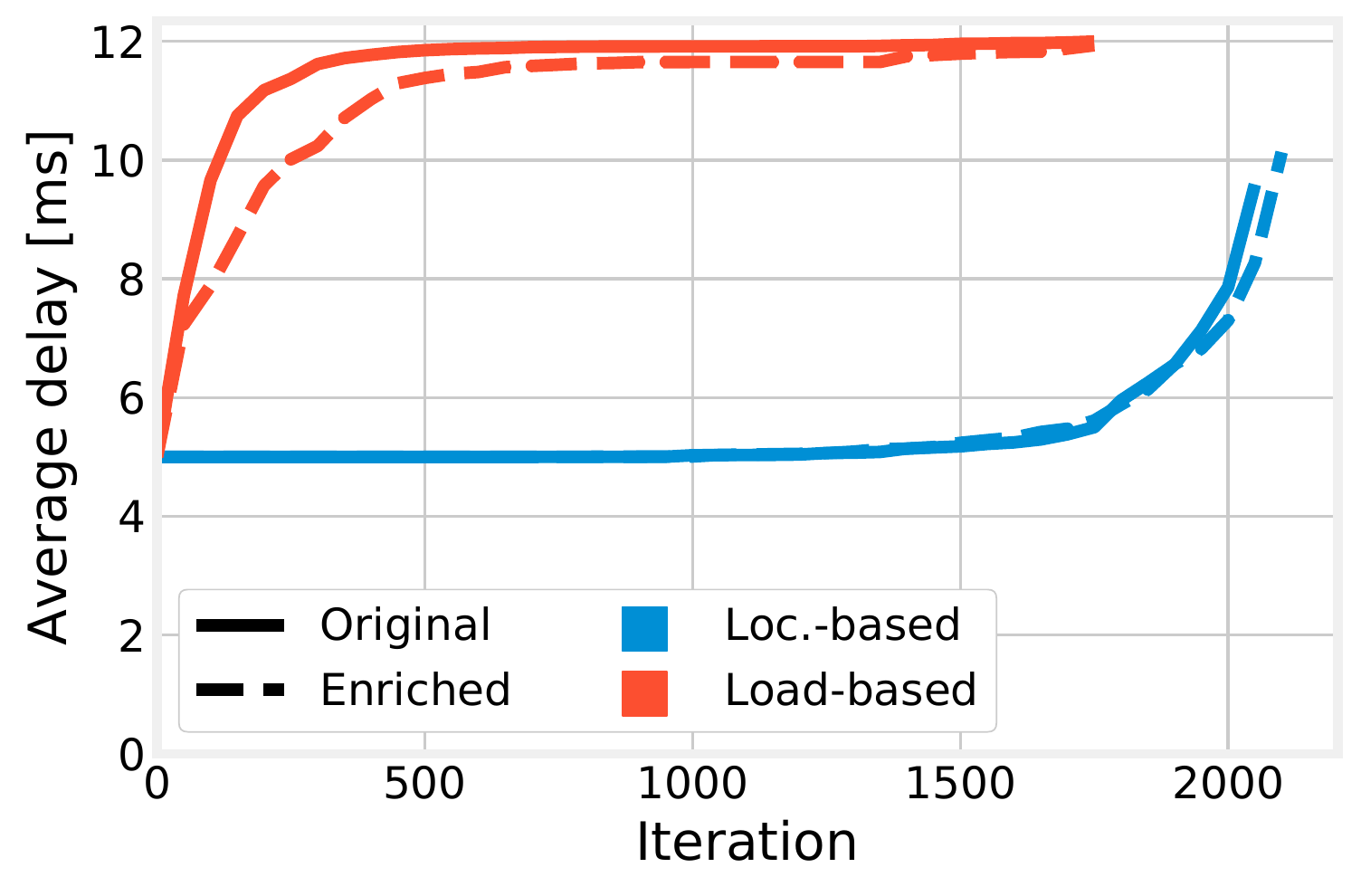}
}
\subfigure[\label{fig:eff}]{
\includegraphics[width=.3\textwidth]{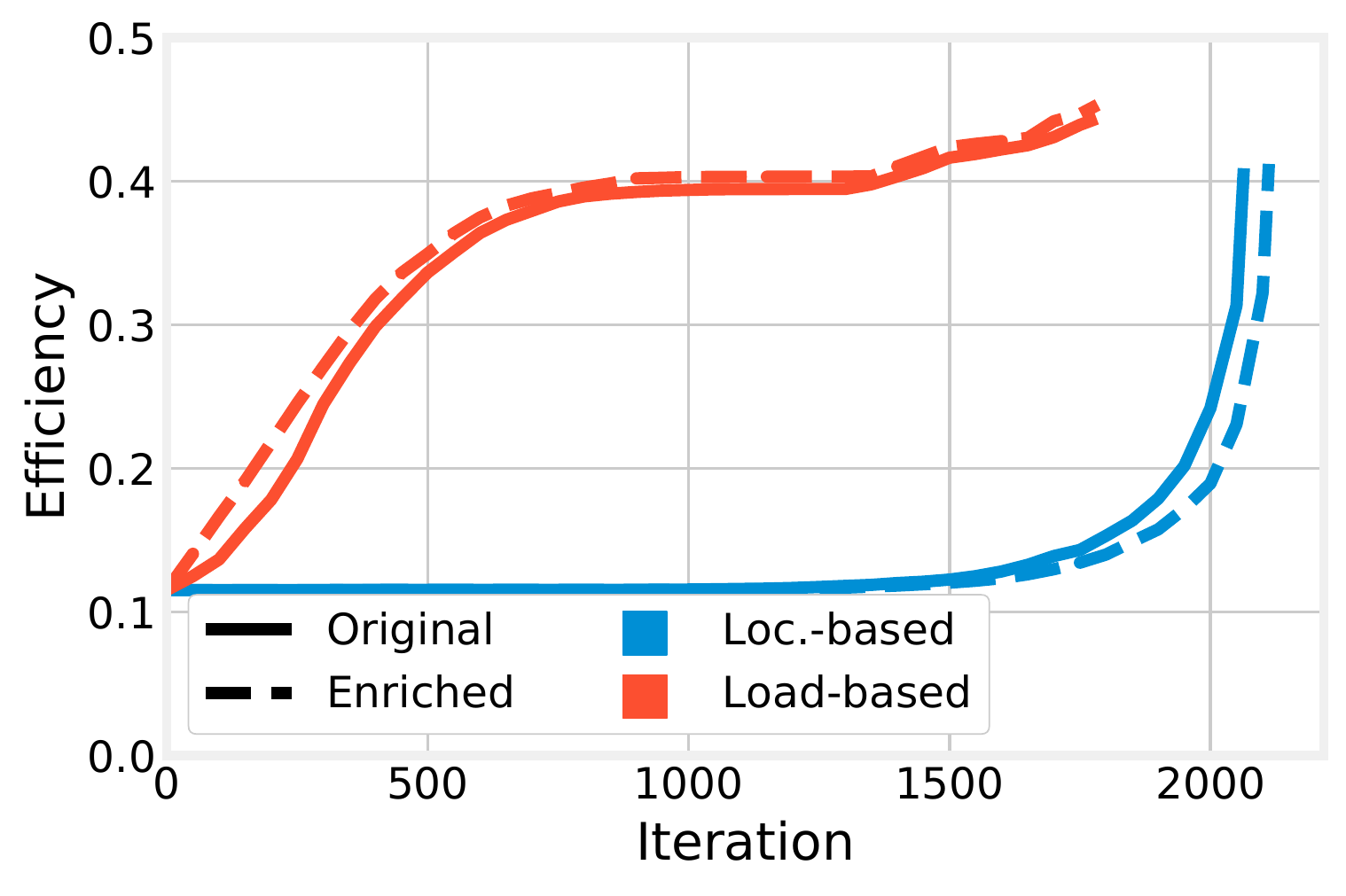}
}
\subfigure[\label{fig:delayeff}]{
\includegraphics[width=.3\textwidth]{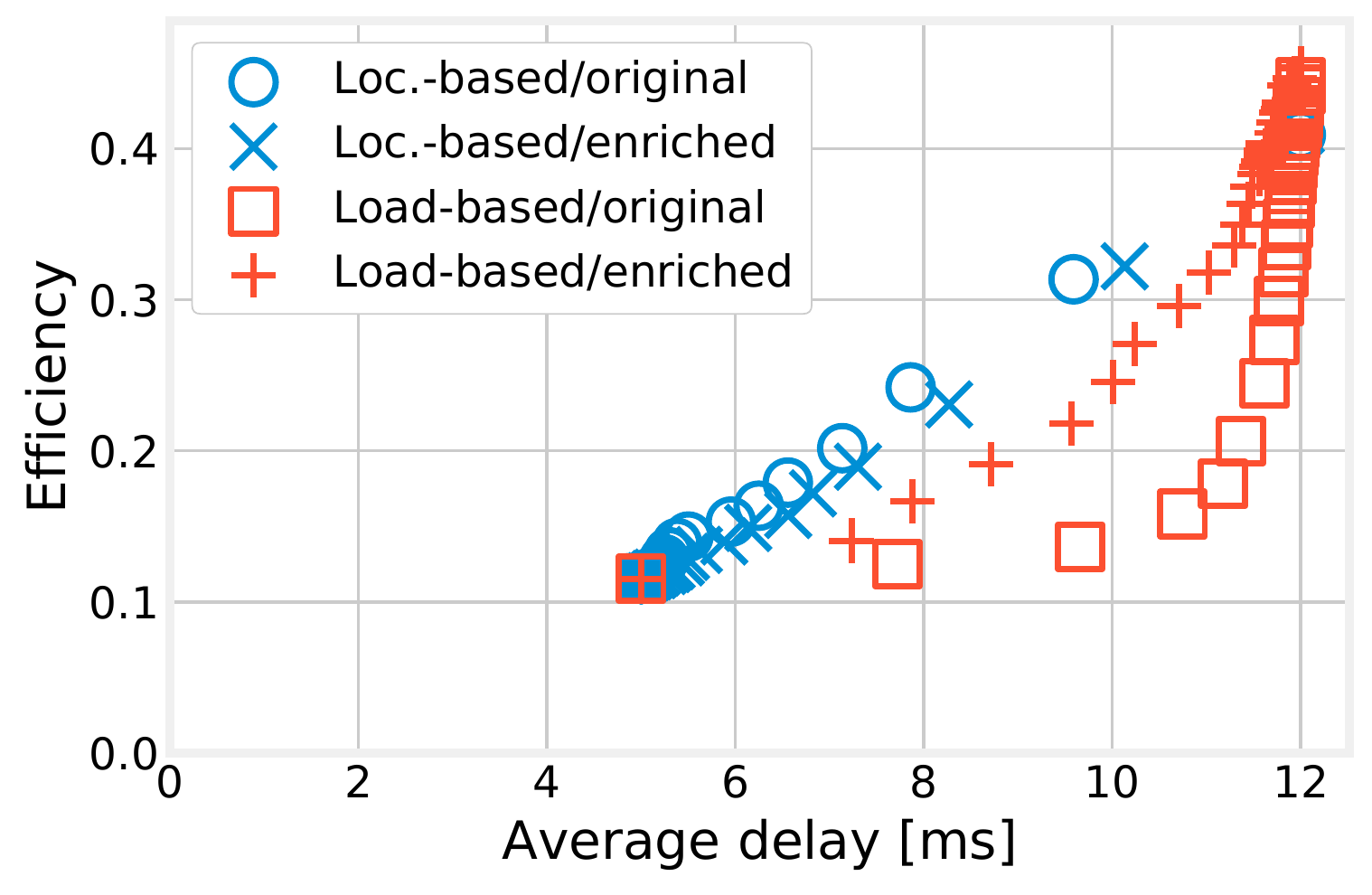}
}
\caption{
Latency (a) and efficiency (b) for each iteration of \Alg{design} for different deployment strategies when the original (solid lines) and enriched (dotted lines) trace is used; resulting latency/efficiency trade-offs (c).
} 
\centering
\subfigure[\label{fig:delay-split}]{
\includegraphics[width=.3\textwidth]{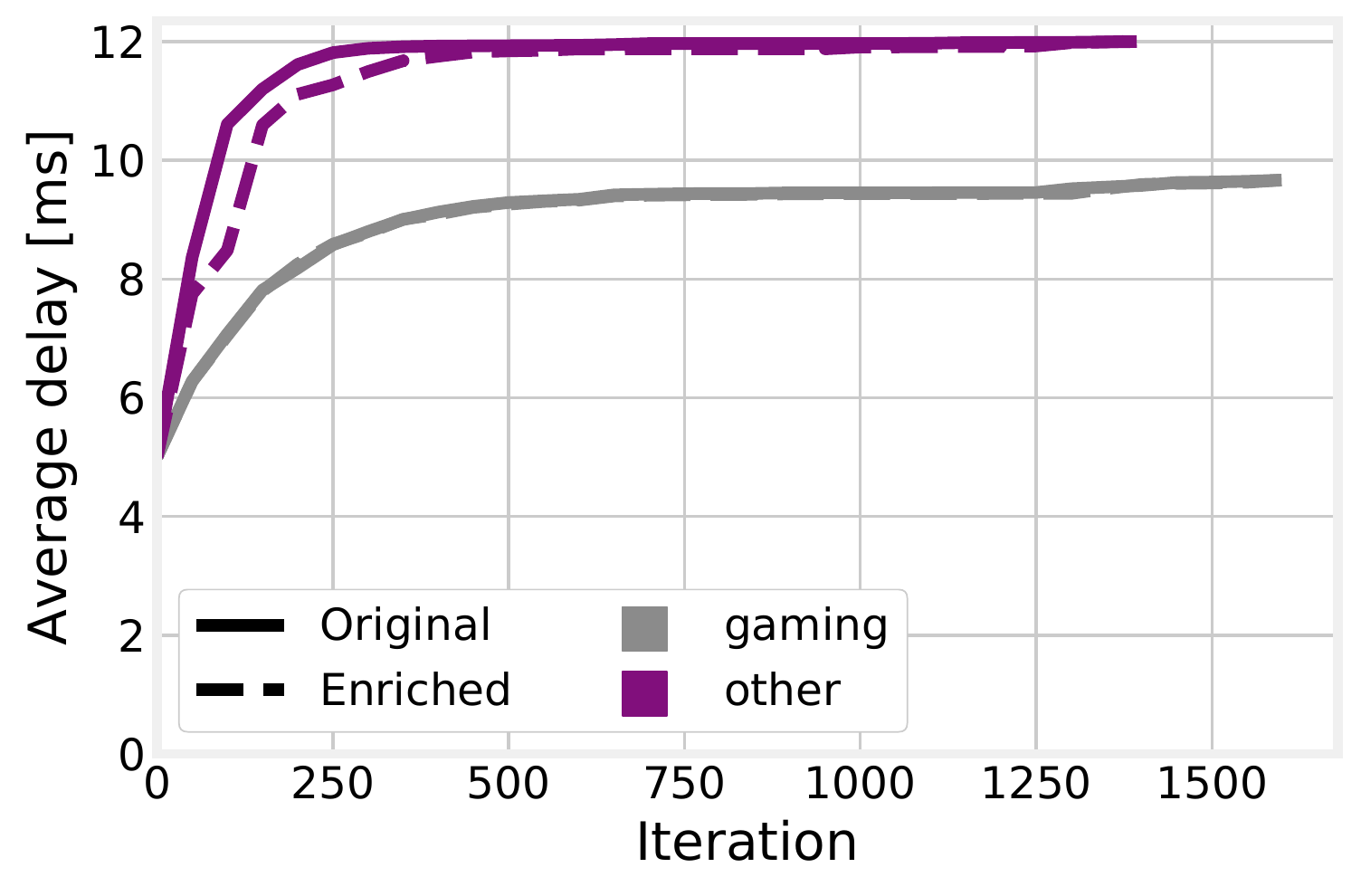}
}
\subfigure[\label{fig:filled-split}]{
\includegraphics[width=.3\textwidth]{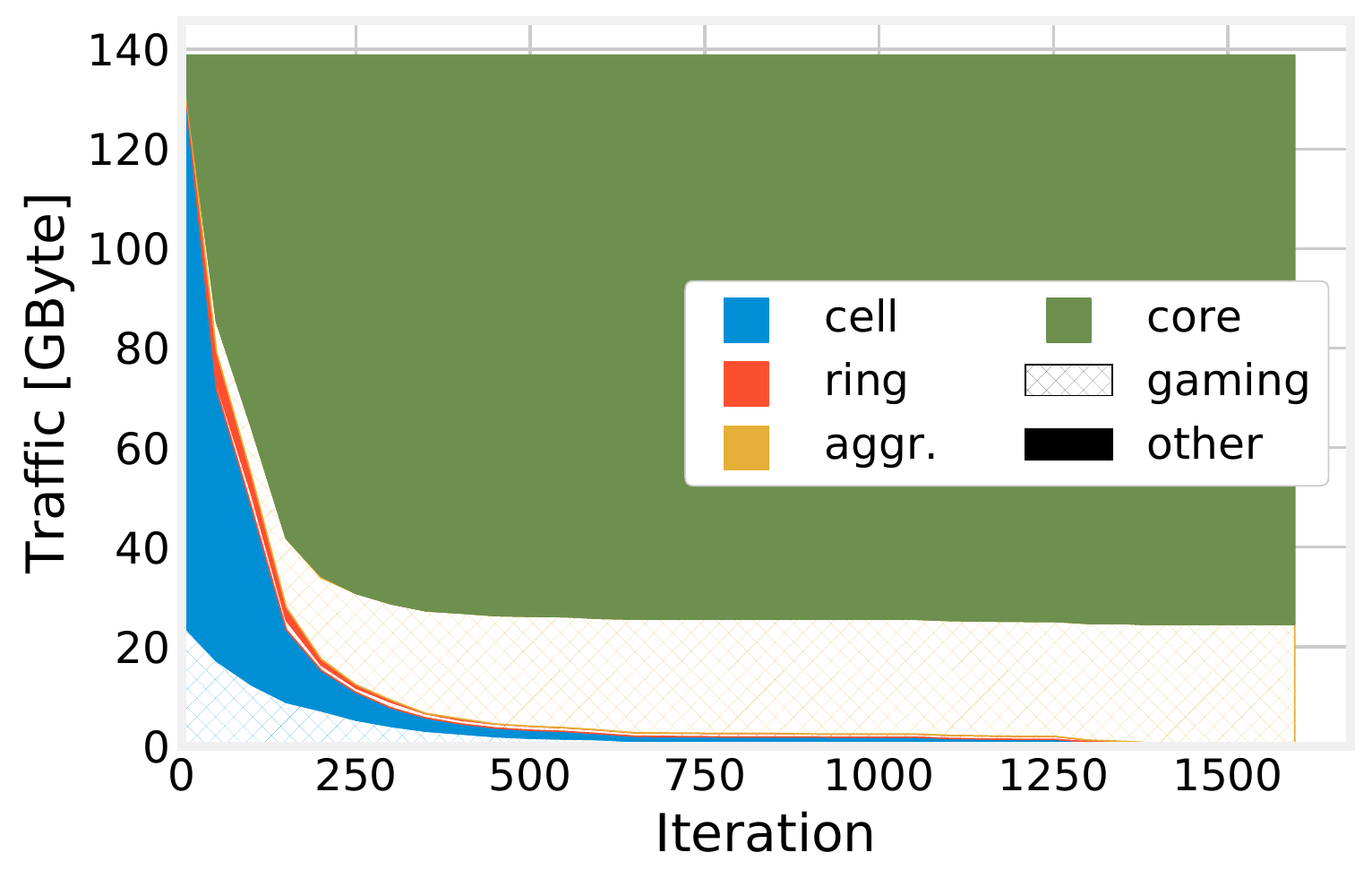}
}
\subfigure[\label{fig:eff-split}]{
\includegraphics[width=.3\textwidth]{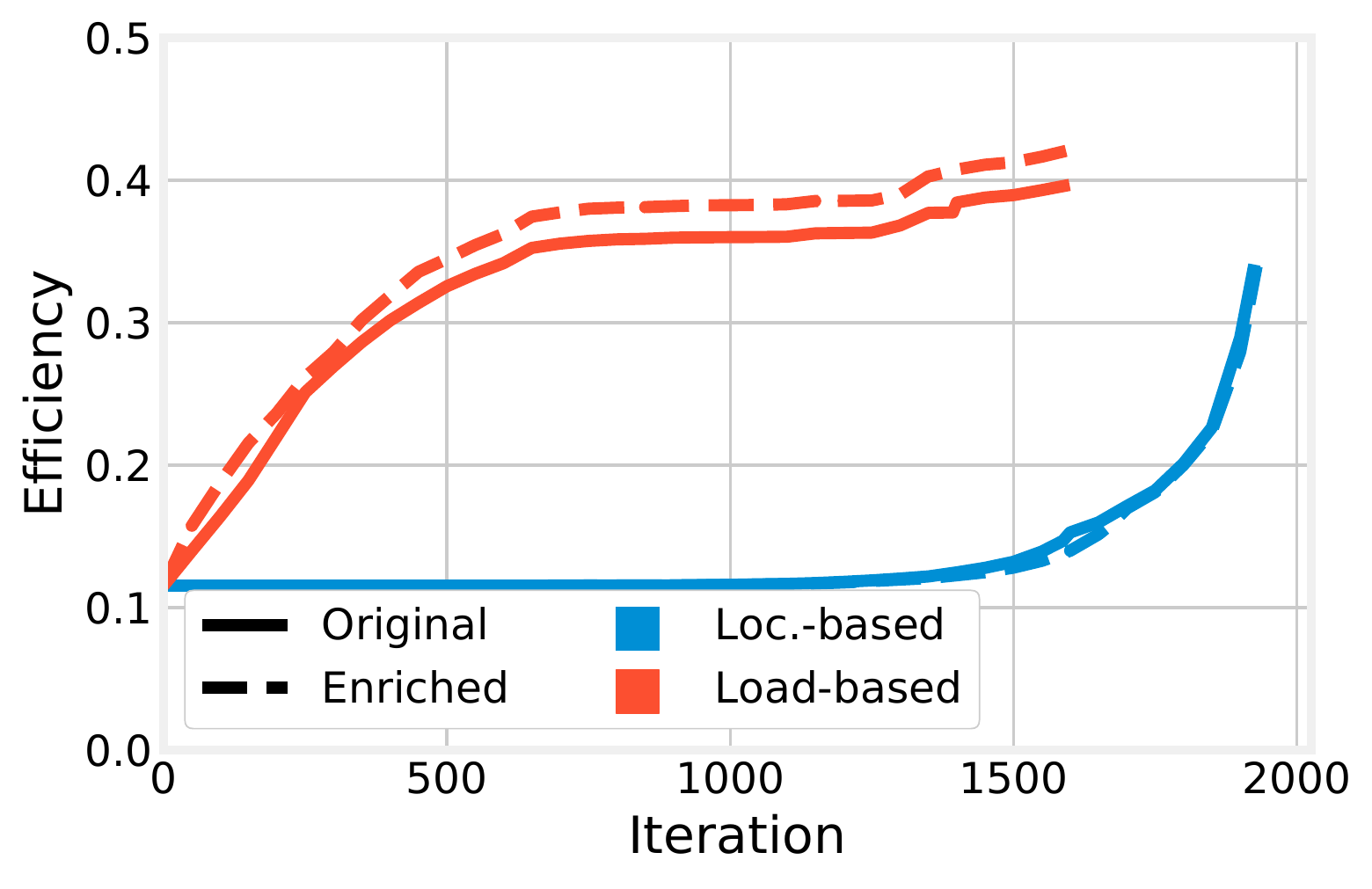}
}
\caption{
Finite latency limit for gaming content: latency for each type of content with load-based scores (a); where different types of content are processed in case of load-based scores (b); resulting efficiency (c).
} 
\end{figure*}

\subsubsection{Multiple categories}
\label{sec:multi}

As mentioned in \Sec{algos}, both \Alg{design} and \Alg{consolidate} make decisions on a {\em per-category} basis, and are therefore able to reproduce the fact that different applications can require different deployments. In the following, we
consider the presence of different application categories and describe how our approach can be easily
leveraged to realize an application-aware deployment. We proceed as follows:
\begin{itemize}
\item we divide application into categories;
\item we run our greedy deployment procedure separately for each category;
\item we combine the resulting deployments.
\end{itemize}
Running \Alg{design} separately for different categories also means evaluating \Line{while} therein using the maximum latency values~$L_{\max}$ of each category. Similarly, the demand-aware distance \Eq{dist-load} is computed separately for each category, only accounting for the contents falling in the current category.

\section{Numerical results}
\label{sec:results}

In this section, we investigate how deployment strategies and score definitions (\Sec{sub-strategy}), using enriched traces (\Sec{sub-ticks}) and considering per-category latency limits (\Sec{sub-latency}), impact the resulting MEC deployment and its effectiveness. For sake of brevity, we only present results for one of the three operators included in our trace. High-resolution versions of the plots for all operators, as well as the Matplotlib source code to generate them, are available from~\cite{html}.

\subsection{Effect of the deployment strategy}
\label{sec:sub-strategy}

The first aspect we are interested in is the impact of the
score definition we adopt, i.e., whether we use \Eq{dist-dist} or \Eq{dist-load} to
implement the \path{score} function in \Alg{design}. To this end, we first assume no latency limit, i.e., $L_{\max}=\infty$, and study (i) how much traffic is processed at each level of the network -- base station (BS), ring, aggregation, core -- and (ii) how many servers are deployed therein.

\Fig{fillnsrv-loc} and \Fig{fillnsrv-dem} demonstrate how \Alg{design} and the consolidation procedure work
in the case of location-based and load-based scores, respectively. We start at iteration~$0$ with one server at each BS; then, at each iteration, we reduce the total number of servers through consolidation, replacing BS servers with servers placed at the higher levels of the network topology. We can  observe that, while \Fig{fillnsrv-loc}(left) shows a significant number of servers at ring and aggregation nodes, \Fig{fillnsrv-dem}(right) shows a tendency to process traffic either at BSs, or at very few, core-level servers.
This is confirmed by \Fig{fillnsrv-loc}(right) and \Fig{fillnsrv-dem}(right): with location-based scores, servers are {\em lazily} moved from BSs to rings, and then further up, only when required. On the other hand, using load-based scores we aggressively process as much data as possible as high in the topology as possible, so as to smoothen the peaks.

\subsection{Efficiency and latency: the importance of ticks}
\label{sec:sub-ticks}

As we have seen, different deployment strategies (i.e., different scores) result in deeply different network planning decisions. In the following, we study (i) how such decisions impact the latency and efficiency of the resulting network, and (ii) if enriching the trace as discussed in \Sec{enrich} impacts either metric.

We estimate the {\em latency} using~\cite{ref-delay} as a reference: the connection between users (UEs in LTE terminology) and base stations (eNBs) requires 5~ms, while additional hops within the backhaul network (e.g., from BSs to rings) require about 2.3~ms each.

As for {\em efficiency}, we define it as the ratio between the average capacity that is actually required to process the traffic and the total capacity deployed throughout the network:
\begin{equation}
\label{eq:eff}
\eta=\frac{
\frac{1}{|\Tc|}\sum_{t\in\Tc}\sum_{n\in\Nc}\sum_{k\in\Kc}\tau_k s(n)^k_t
}{
\sum_{n\in\Nc}\max_{t\in\Tc}\sum_{k\in\Kc} \tau_k s(n)^k_t
}.
\end{equation}
A network where all servers are always fully utilized would have an efficiency of~$1$, while a network where servers are almost never used would have almost-zero efficiency.
Also notice that in \Eq{eff} we consider all content categories, and abuse the notation to indicate with~$s(n)^k_t$
the amount of data of content of category~$k$ that node~$n$ must process at time~$t$.

\Fig{delay} shows that location-based scores are associated with a lower latency than load-based ones. On the other hand, as shown in \Fig{eff}, load-based scores maximize the network efficiency -- which, however, hardly exceeds 30\%, due to time fluctuations of the demand. It is interesting to compare the solid and dashed lines in \Fig{delay} and \Fig{eff}, respectively obtained with the original trace and the enriched trace. Using the enriched trace, i.e., ticks instead of megabytes, provides both a lower latency {\em and} a higher efficiency. This is confirmed by \Fig{delayeff}, depicting the latency/efficiency trade-offs that can be obtained using different scores and traces.

The difference between location- and load-based performance in \Fig{delay}--\Fig{delayeff} can be interpreted as the benefit obtained by using the enriched trace {\em in lieu} of the original one. By doing so, we can make better deployment decisions, which consistently result in better efficiency and lower latency.

\subsection{The effect of latency limits}
\label{sec:sub-latency}

We now consider finite latency limits; specifically, we set~$L_{\max}=50$~ms for video and maps and,
based on~\cite{ref-maxdelay}, $L_{\max}=10$~ms for gaming traffic. \Fig{delay-split}, obtained with load-based scores, shows that different content categories now experience different latencies. In particular, gaming data are always served within its maximum latency limit~$L_{\max}=10$~ms, while other content is served with a higher latency --
but still lower than its maximum limit. This is due to the different locations within the network where content is processed, as shown in \Fig{filled-split}. In the late iterations of the algorithm, we can observe that almost all gaming content (represented by patterned areas) is served at aggregation nodes, while almost all non-gaming content (solid areas) is served at core nodes. Indeed, based on~\cite{ref-delay}, going from UEs to core nodes entails a latency of 12~ms, which is not compatible with the latency limit for gaming.
Furthermore, it is important to note that, even if latency requirements for video and maps would permit the deployment of their servers in the cloud, MEC is still an appealing solution as it avoids  transferring large amounts of data over long distances, thus reducing bandwidth consumption.

Being unable to serve gaming content at core servers also has the potential to impair network efficiency; indeed, as it can be seen from \Fig{eff} and \Fig{delayeff}, the highest efficiency is reached when most content is processed there. By comparing \Fig{eff-split} to \Fig{eff}, we can indeed observe a decrease in the efficiency; however, such a decrease is lower than 10\%, which confirms that MEC is able to deliver both low latency and high efficiency.

Finally, it is interesting to observe that the advantage of using enriched traces {\em in lieu} of the original is clearly visible, in terms of both latency and efficiency.

\section{Conclusion and future work}
\label{sec:conclusion}

One of the challenges we face in the design of next-generation networks is using real-world data about {\em present-day} technologies to predict their performance. In this work, we took cellular networks as a case study, and demonstrated how a real-world, large-scale dataset about data demand in LTE can be leveraged to design a next-generation, MEC-based network.

To this end, we {\em enriched} the dataset at our disposal, integrating it with information about the processing power needed to generate traffic of the two main categories of content. We obtained such information through a set of hands-on experiments, based on the Minecraft and FFserver open-source servers. Our results show that using the enriched trace instead of the original one results in better network design, allowing both lower latency {\em and} higher network efficiency.

\bibliographystyle{IEEEtran}
\bibliography{refs}

\begin{IEEEbiography}[{\includegraphics[width=1in,height=1.25in,clip,keepaspectratio]{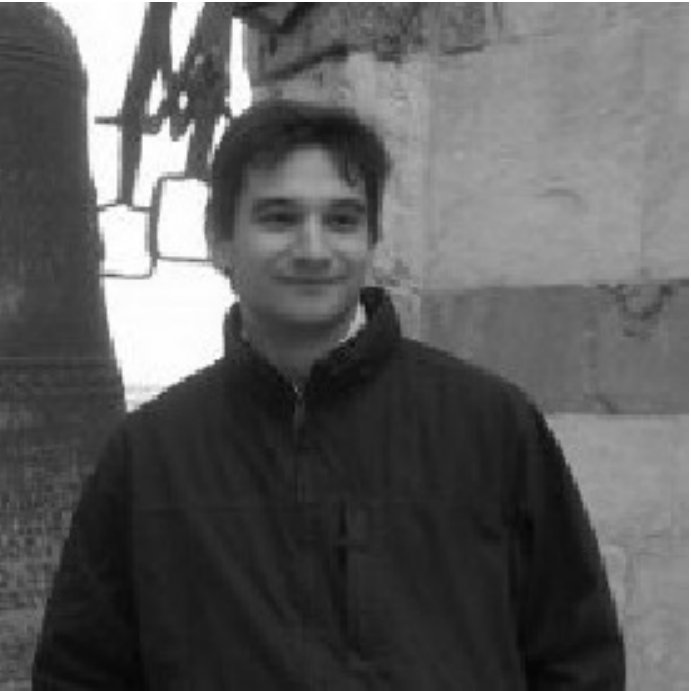}}]%
{Francesco Malandrino} (M'15) earned his Ph.D. in 2012 from Politecnico di Torino, Italy, where he is currently an assistant professor. Before his current appointment, he held short-term positions at Trinity College, Dublin, and at the Hebrew University of Jerusalem as a Fibonacci Fellow. His interests focus on wireless and vehicular networks and infrastructure management.
\end{IEEEbiography}

\begin{IEEEbiography}[{\includegraphics[width=1in,height=1.25in,clip,keepaspectratio]{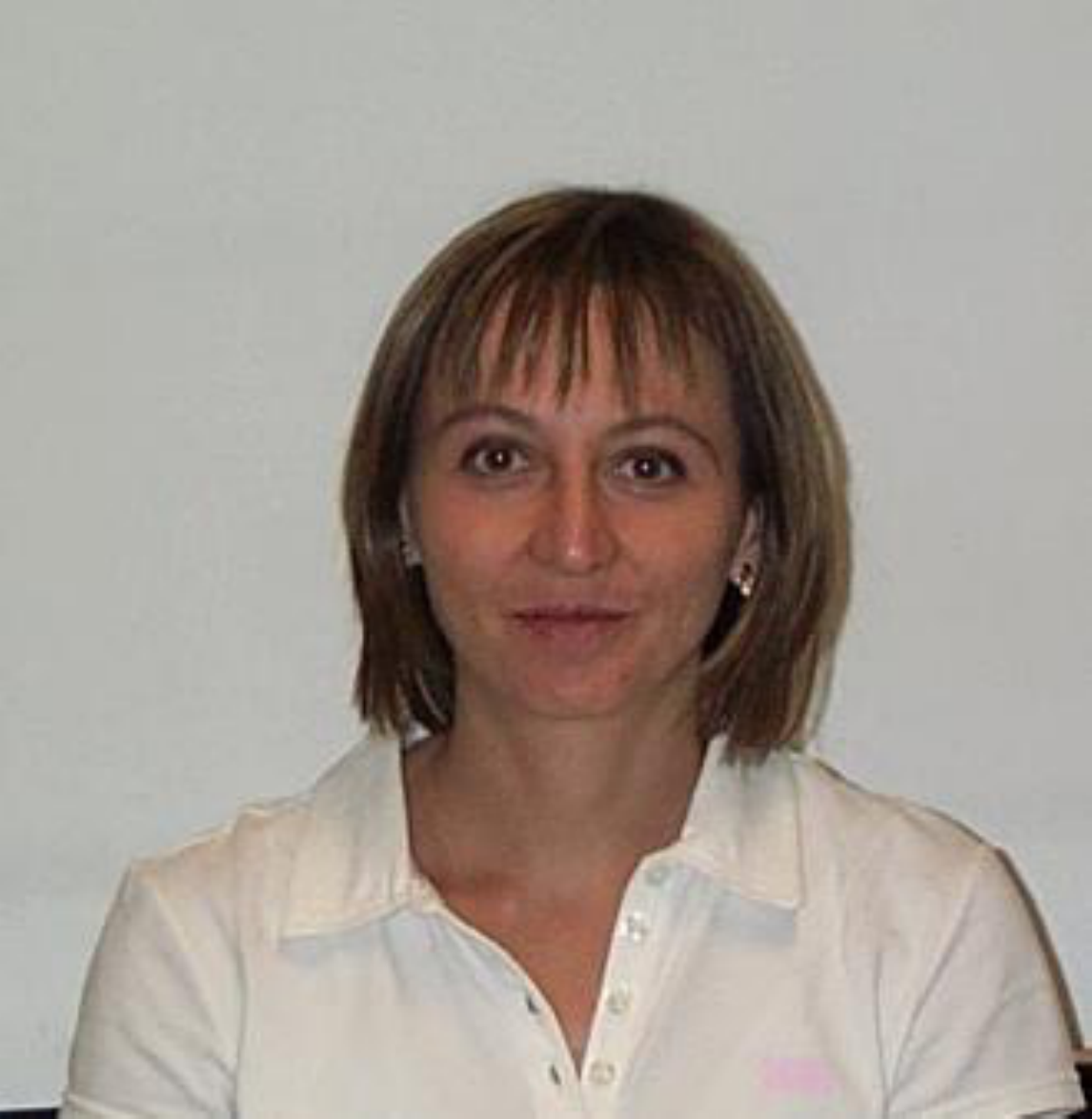}}]%
{Carla-Fabiana Chiasserini} (M'98, SM'09, F'18) graduated in Electrical Engineering (summa cum laude) from
the University of Florence in 1996. She received her Ph.D. from Politecnico di Torino, Italy, in
2000.  She has worked as a visiting researcher at UCSD in 1998--2003, and she is currently an
Associate  Professor  with  the  Department  of Electronic  Engineering  and  Telecommunications  at
Politecnico di Torino.  Her research interests include architectures, protocols, and performance
analysis of wireless networks. Dr. Chiasserini has published over 230 papers in prestigious journals
and leading international conferences, and she serves as Associated Editor of several journals.
\end{IEEEbiography}

\begin{IEEEbiography}[{\includegraphics[width=1in,height=1.25in,clip,keepaspectratio]{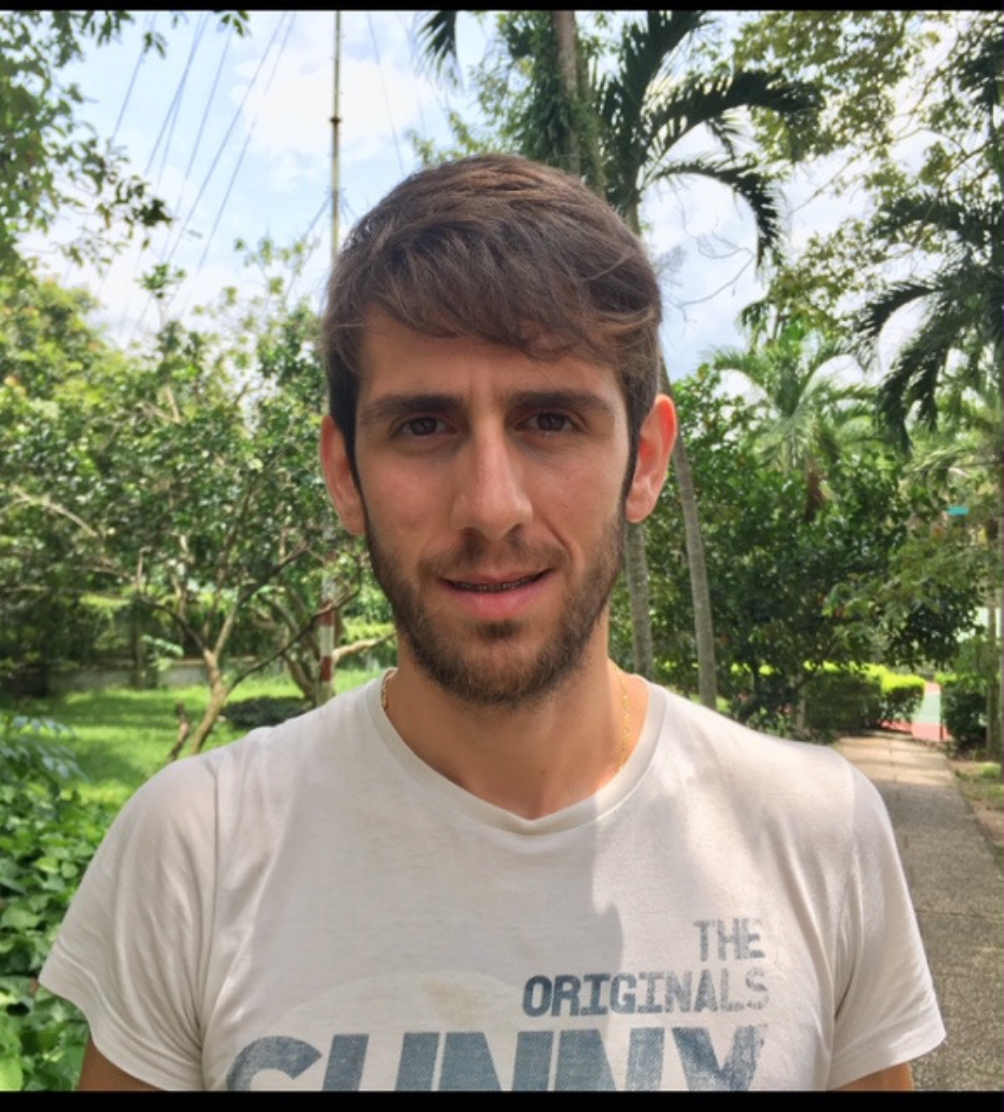}}]%
{Giuseppe Avino} obtained his master degree in Computer and Communication Network Engineering in 2016 from Politecnico di Torino. From January to October 2017, he carried out research on container virtualization at Politecnico di Torino, where he started his Ph.D. in November 2017. His interests focus on wireless network communication and container virtualization.
\end{IEEEbiography}

\begin{IEEEbiography}[{\includegraphics[width=1in,height=1.25in,clip,keepaspectratio]{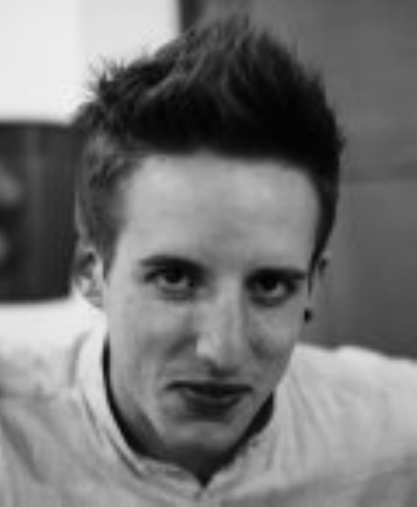}}]%
{Marco Malinverno} is a Ph.D. student at Politecnico di Torino. He obtained his master degree in Computer and Communication Networks Engineering in 2016. His research activities range from containerization techniques to vehicular wireless networks.
\end{IEEEbiography}

\begin{IEEEbiography}[{\includegraphics[width=1in,height=1.25in,clip,keepaspectratio]{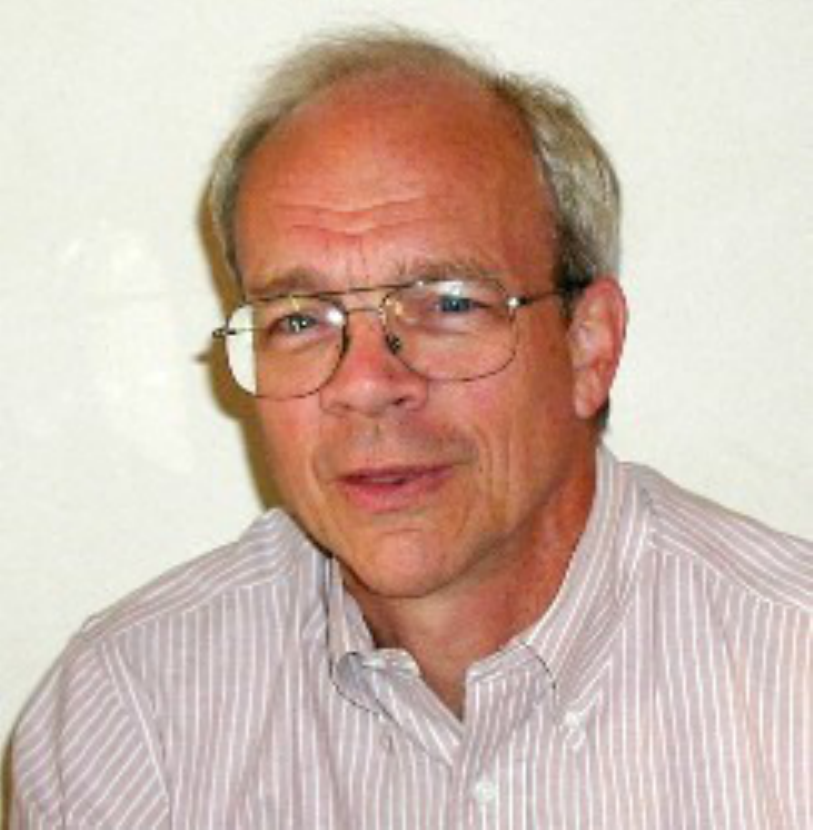}}]%
{Scott Kirkpatrick} (SM '80, F '91) has been a professor in the computer science department at the Hebrew University, in Jerusalem, Israel, since 2000. Before that he was at IBM Research, Yorktown Heights, where he did research in physics and engineering, developing simulated annealing and IBM’s first tablet computers. Professor Kirkpatrick is a Fellow of the APS, the AAAS, the IEEE, and the ACM, has written over 100 papers and holds more than 10 patents. He holds an AB from Princeton University and PhD (in physics) from Harvard University.
\end{IEEEbiography}

\end{document}